\documentclass{article}
\usepackage[dvips]{graphicx}
\usepackage{amsmath}
\usepackage{amssymb}
\usepackage{amsfonts}
\usepackage{bm}

\begin{document}



\title{\bf Linear Response Theory of Scale-Dependent \\
Viscoelasticity for Overdamped Brownian \\
Particle Systems}
\author{Takashi Uneyama\\
\\
JST-PRESTO, and 
Department of Materials Physics, \\
Graduate School of Engineering,  Nagoya University,\\
Furo-cho, Chikusa, Nagoya 464-8603, Japan}
\date{}


\maketitle

\begin{abstract}
We show the linear response theory of spatial-scale-dependent 
relaxation moduli for overdamped Brownian particle systems.
We employ the Irving-Kirkwood stress tensor field as the microscopic
stress tensor field. We show that the scale-dependent relaxation modulus tensor,
which characterizes the response of the stress tensor field to the 
applied velocity gradient field, can be expressed by using the
correlation function of the Irving-Kirkwood stress tensor field.
The spatial Fourier transform of the relaxation modulus tensor
gives the wavenumber-dependent relaxation modulus.
For isotropic and homogeneous systems, the relaxation modulus tensor
has only two independent components. The transverse and longitudinal
deformation modes give the wavenumber-dependent shear relaxation modulus and
the wavenumber-dependent bulk relaxation modulus.
As simple examples, we derive the explicit expressions for the relaxation moduli for
two simple models the non-interacting Brownian particles and
the harmonic dumbbell model.
\end{abstract}

%

\section{INTRODUCTION}

To measure rheological properties, (macroscopic) rheometers are widely
employed. We apply the macroscopic strain (or stress) to the sample, and measure
the macroscopic stress (or the strain) as the response. Then we obtain
rheological properties such as the linear viscoelasticity.
In some cases, measurements by rheometers become difficult.
The typical amount of a sample required for the
measurements is about $1 \, \text{ml}$.
If the amount of a sample is limited, the measurements become difficult.
The microrheology measurements have been proposed to
obtain rheological properties with small amount of samples.

In the microrheology, small spherical colloidal particles are dispersed in the
sample and then the dynamics of the colloidal particles is measured\cite{Waigh-2005,Squires-Mason-2009}.
The dynamics of the colloidal particle is assumed to obey the generalized
Langevin equation:
\begin{equation}
 \label{generalized_langevin_equation}
 0 = - \int_{-\infty}^{t} dt' \,  \Gamma(t - t') \frac{d\bm{R}(t')}{dt'}
  + \bm{\Xi}(t),
\end{equation}
where $\bm{R}(t)$ is the center of mass position of the colloid particle
in a three dimensional space,
$\Gamma(t)$ is the friction kernel,
and $\bm{\Xi}(t)$ is the Gaussian colored noise.
(In eq~\eqref{generalized_langevin_equation}, the inertia term
is dropped by assuming that the momentum relaxation is sufficiently fast.)
From the
fluctuation-dissipation relation, the noise $\bm{\Xi}(t)$ satisfies
\begin{equation}
 \label{fluctuation_dissipation_relation_generalized_langevin}
 \langle \bm{\Xi}(t) \rangle = 0, \qquad
 \langle \bm{\Xi}(t) \bm{\Xi}(t') \rangle = k_{B} T \Gamma(|t - t'|)\bm{1}.
\end{equation}
Here, $\langle \dots \rangle$ represents the statistical average,
$k_{B}$ is the Boltzmann constant, $T$ is the temperature, and $\bm{1}$
is the unit tensor.
If we {\em assume} that the sample around the particle behaves as
viscoelastic fluid which has the same linear viscoelasticity at
the macroscopic scale, the friction kernel $\Gamma(t)$ can be
related to the linear viscoelasticity of the sample (the generalized
Stokes relation):
\begin{equation}
 \label{generalized_stokes_relation}
 \Gamma(t) = 6 \pi a G(t).
\end{equation}
$a$ is the radius of the particle and $G(t)$ is the macroscopic
shear relaxation modulus. The mean squared displacement (MSD) of
the particle can be related to the friction kernel\cite{Fox-1977}. By combining
the MSD and eq~\eqref{generalized_stokes_relation}, we have
the so-called the generalized Stokes-Einstein relation\cite{Waigh-2005,Squires-Mason-2009}:
\begin{equation}
 \label{generalized_stokes_einstein_relation}
 \langle [\bm{R}(t) - \bm{R}(0)]^{2} \rangle
  = \frac{k_{B} T}{\pi a} J(t),
\end{equation}
where $J(t)$ is the macroscopic shear creep compliance.
Therefore, we can estimate the linear viscoelasticity of the sample
by measuring the MSD of a colloidal particle.

The microrheological methods explained above sounds interesting.
However, we should point that its validity is not fully guaranteed.
The {\em assumption} that the sample around the particle has the
same linear viscoelasticity as the macroscopic one is generally not
correct. The fact that the viscoelasticity depends on the length
scale (or the wavenumber scale) is well known\cite{Evans-1981,Alley-Alder-1983,Evans-Morris-book,Hansen-Daivis-Travis-Todd-2007,Glavatskiy-Dalton-Daivis-Todd-2015}. Therefore,
the generalized Stokes-Einstein relation \eqref{generalized_stokes_einstein_relation} does {\em not} hold in general.
What we obtain from the microrheology measurements is rather
the spatial-scale-dependent linear viscoelasticity than the macroscopic linear viscoelasticity.
(In what follows, we call the spatial-scale-dependent linear viscoelasticity
as the spatial-dependent linear viscoelasticity.)
We will need a relation between the MSD and the scale-dependent linear
viscoelasticity, instead of eq~\eqref{generalized_stokes_einstein_relation}.
However, the flow field around a colloidal particle is not simple and
the relation between the MSD and the scale-dependent linear viscoelasticity
will not be simple.
The scale-dependent linear viscoelasticity itself is not well understood
compared with the macroscopic linear viscoelasticity.
Therefore, to correctly understand the microrheology measurements and utilize them
to study various rheological properties, we need the microscopic theory
of scale-dependent linear viscoelasticity as a first step.

The scale-dependent linear viscoelasticity will be also useful to
study spatially inhomogeneous systems. The macroscopic linear viscoelasticity
reflects the relaxation dynamics of molecules in the sample.
Naively, we expect that the relaxation depends both on the temporal and
spatial scales.
The temporal scale of the relaxation can be evaluated by the macroscopic
linear viscoelasticity.
However, the information on
the spatial scale of the relaxation cannot be obtained from the macroscopic
linear viscoelasticity.
The use of the scale-dependent linear viscoelasticity
will enable us to extract the characteristic
spatial scale of the relaxation dynamics.

Some researchers studied the scale-dependent linear viscoelasticity by molecular
dynamics simulations and calculated several quantities such as the
wavenumber-dependent shear viscosity\cite{Evans-1981,Alley-Alder-1983,Evans-Morris-book,Hansen-Daivis-Travis-Todd-2007,Glavatskiy-Dalton-Daivis-Todd-2015}.
As far as the author knows, however, most of studies are based on the
Hamiltonian dynamics. In the field of rheology, the overdamped
Langevin dynamics is widely utilized as the microscopic model.
For example, polymer dynamics models such as the Rouse model and the reptation
model are based on the overdamped Langevin equation\cite{Doi-Edwards-book}.
These models can reasonably explain the rheological properties which
are measured by the (macroscopic) rheometers.
The theory of scale-dependent linear viscoelasticity based on
the overdamped Langevin equation will be required to study
the microrheology or the spatial scale of the relaxation.

To theoretically analyze the scale-dependent linear viscoelasticity,
we need the expression of the position-dependent stress tensor field.
As a position-dependent stress tensor field,
so-called the Irving-Kirkwood stress\cite{Irving-Kirkwood-1950,Schfield-Henderson-1982}
is widely utilized.
The Irving-Kirkwood stress tensor field is derived on the basis of the
conservation equation for the momentum field.
In the linear response theory, the stress field should be defined as the
thermodynamic conjugate to the applied strain field.
Therefore, whether the linear response is actually
described by the auto correlation function of the Irving-Kirkwood stress
tensor field or not is not fully clear.

In this work, we consider the linear response theory for the microscopic
stress field to the applied microscopic strain field.
We consider the interacting Brownian particle systems, which obey
the overdamped Langevin equations, as the target system.
We assume that the microscopic deformation field and
the microscopic velocity field for the target system can be controlled.
This is consistent with the Langevin equation with the external flow
field, which is often employed to analyze and simulate the dynamics
of soft matters such as polymers and colloids.
We define the microscopic stress tensor field and the velocity
gradient tensor field, and construct the linear response theory.
We show that the Irving-Kirkwood stress tensor field can be successfully
employed as the microscopic stress tensor field, and the relaxation modulus
tensor is given as the equilibrium correlation function of
the microscopic stress tensor field.
As simple examples, we analytically calculate relaxation moduli
for the non-interacting Brownian particles and the harmonic
dumbbell model.

\section{THEORY}

\subsection{Microscopic Strain Tensor and Microscopic Stress Tensor}

We consider overdamped Brownian particle systems where we do not
have the degrees of freedom of momenta.
For simplicity, we assume that our system is statistically isotropic and
homogeneous.
(The statistical properties of the system is not changed under
the rotation and the translation.)
We assume that the
particles interact by pairwise potentials. 
We describe the
position of the $i$-th particle as $\bm{R}_{i}$. The potential energy of the
system is
\begin{equation}
 \mathcal{U}(\lbrace \bm{R}_{i} \rbrace) =
  \sum_{i > j} \phi_{ij}(R_{ij}),
\end{equation}
where $ R_{ij} \equiv |\bm{R}_{ij}|$ and
$\bm{R}_{ij} \equiv \bm{R}_{i} - \bm{R}_{j}$, and $\phi_{ij}(r)$ is the interaction
potential between particles $i$ and $j$. We assume that the interaction potential
depends only on the distance between two particles.

First we consider to impose a deformation to the system. The
deformation is characterized by the displacement field $\bm{u}(\bm{r})$.
The particle positions after the deformation is
\begin{equation}
 \bm{R}_{i}' \equiv \bm{R}_{i} + \bm{u}(\bm{R}_{i}).
\end{equation}
We assume that the displacement field is smooth.
Then the change of the potential energy by the deformation can
be expanded into the series 
\begin{equation}
 \label{potential_energy_difference}
 \mathcal{U}(\lbrace \bm{R}_{i}' \rbrace)
  - \mathcal{U}(\lbrace \bm{R}_{i} \rbrace)
  = \sum_{i > j} \frac{\partial \mathcal{U}(\lbrace \bm{R}_{i} \rbrace)}
  {\partial \bm{R}_{ij}} \cdot [\bm{u}(\bm{R}_{i}) - \bm{u}(\bm{R}_{j}) ] 
  + (\text{higher order terms}).
\end{equation}
The higher order terms in eq~\eqref{potential_energy_difference} can be safely neglected when the relative
displacements of particles are small. This situation can be realized when
$\nabla \bm{u}(\bm{r})$ is small. We assume that $\nabla \bm{u}(\bm{r})$ is
sufficiently small, and simply drop higher order terms in eq~\eqref{potential_energy_difference}
in what follows.
From the operational point of view, we may interpret that the system is deformed by the externally
applied strain field $\bm{\epsilon}(\bm{r})$. Then the change of the potential energy should be
expressed by using the strain field and the stress field which is conjugate to the strain field as
\begin{equation}
 \label{potential_energy_difference_target}
 \mathcal{U}(\lbrace \bm{R}_{i}' \rbrace)
  - \mathcal{U}(\lbrace \bm{R}_{i} \rbrace)
  = \int d\bm{r} \, \hat{\bm{\sigma}}^{(\mathrm{p})}(\bm{r}) :
  \bm{\epsilon}(\bm{r}),
\end{equation}
where $\hat{\bm{\sigma}}^{(\mathrm{p})}(\bm{r})$ is
the microscopic stress tensor field which is related to the potential (the potential part).
We assume that the strain field can be simply related to the displacement field $\bm{u}(\bm{r})$ as\cite{Landau-Lifshitz-book}
\begin{equation}
 \bm{\epsilon}(\bm{r}) = \frac{1}{2}
  \left[ \nabla \bm{u}(\bm{r}) + [\nabla \bm{u}(\bm{r})]^{\mathrm{T}} \right].
\end{equation}

\begin{figure}[tb]
 \centering
 \includegraphics[width=0.3\linewidth,clip]{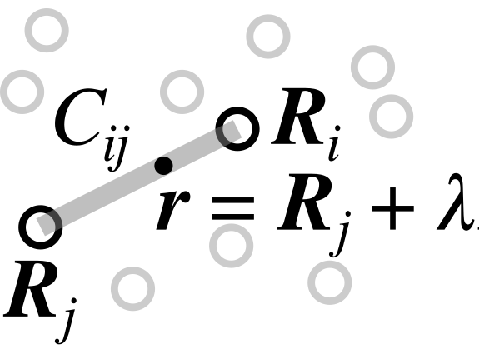}
\caption{\label{irving_kirkwood_path}
 The system consists of many particles (expressed by circles).
 The path of integration $C_{ij}$ which connect the $i$-th and $j$-th particles (black circles)
 is required.
 The straight path from the $j$-th particle (at $\bm{R}_{j}$) to the $i$-th particle (at $\bm{R}_{i}$)
 is employed (the thick gray line, the Irving-Kirkwood path). The point on the path (the black dot) can be expressed as $\bm{r} = \bm{R}_{j} + \lambda \bm{R}_{ij} $
 with $0 \le \lambda \le 1$.}
\end{figure}

Then our problem is to find the explicit expression of the stress tensor $\hat{\bm{\sigma}}^{(\mathrm{p})}(\bm{r})$
which gives the potential energy change by eq~\eqref{potential_energy_difference}
correctly. Eq~\eqref{potential_energy_difference} can be
rewritten as
\begin{equation}
 \label{potential_energy_difference_modified}
 \begin{split}
  \mathcal{U}(\lbrace \bm{R}_{i}' \rbrace)
  - \mathcal{U}(\lbrace \bm{R}_{i} \rbrace)
  & = \sum_{i > j} \frac{\partial \phi_{ij}(R_{ij})}
  {\partial R_{ij}} \frac{\bm{R}_{ij}}{R_{ij}} [\bm{u}(\bm{R}_{i})
  - \bm{u}(\bm{R}_{j})] \\
  & = \sum_{i > j} \frac{\partial \phi_{ij}(R_{ij})}
  {\partial R_{ij}} \frac{\bm{R}_{ij}}{R_{ij}} \int_{C_{ij}} d\bm{l} \cdot
  \frac{\partial}{\partial \bm{l}} \bm{u}(\bm{l}) ,
 \end{split}
\end{equation}
where the integral in the last line of eq~\eqref{potential_energy_difference_modified} is taken along a path $C_{ij}$,
which starts at $\bm{R}_{j}$ and ends at $\bm{R}_{i}$. Following
the Irving-Kirkwood theory\cite{Irving-Kirkwood-1950,Schfield-Henderson-1982}, we employ the straight path which connects
$\bm{R}_{j}$ and $\bm{R}_{i}$ (the Irving-Kirkwood path)
as $C_{ij}$. The Irving-Kirkwood path is shown in Figure~\ref{irving_kirkwood_path}.
Although the Irving-Kirkwood path may not be a unique candidate,
the Irving-Kirkwood path makes the stress tensor symmetric.
Since the macroscopic stress tensor is symmetric, the symmetric
microscopic stress tensor would be preferred than non-symmetric one.
Thus we employ the Irving-Kirkwood path as $C_{ij}$.
With the Irving-Kirkwood path, eq~\eqref{potential_energy_difference_modified} can be
further rewritten as
\begin{equation}
 \label{potential_energy_difference_modified2}
 \begin{split}
  \mathcal{U}(\lbrace \bm{R}_{i}' \rbrace)
  - \mathcal{U}(\lbrace \bm{R}_{i} \rbrace)
  & = \sum_{i > j} \frac{\partial \phi_{ij}(R_{ij})}
  {\partial R_{ij}} \frac{\bm{R}_{ij}}{R_{ij}} \int_{0}^{1} d \lambda \,
  \bm{R}_{ij} \cdot
  \left. \nabla \bm{u}(\bm{r})
  \right|_{\bm{r} = \bm{R}_{j} + \lambda \bm{R}_{ij}} \\
  & = \int d\bm{r} \sum_{i > j} \frac{\partial \phi_{ij}(R_{ij})}
  {\partial R_{ij}} \frac{\bm{R}_{ij} \bm{R}_{ij} }{R_{ij}} \int_{0}^{1} d \lambda \,
  \delta(\bm{r} - \bm{R}_{j} - \lambda \bm{R}_{ij}) : \nabla \bm{u}(\bm{r}).
 \end{split}
\end{equation}
We compare eqs~\eqref{potential_energy_difference_target}
and \eqref{potential_energy_difference_modified2} to find the expression
for the stress tensor. In eq~\eqref{potential_energy_difference_modified2},
we do not have the strain field $\bm{\epsilon}(\bm{r}) = [\nabla \bm{u}(\bm{r}) + [\nabla \bm{u}(\bm{r})]^{\mathrm{T}}] / 2$
but only have $\nabla \bm{u}(\bm{r})$. Thus we cannot identify which part of eq~\eqref{potential_energy_difference_modified2}
corresponds to the stress tensor.
Because the second rank tensor $\bm{R}_{ij} \bm{R}_{ij}$ 
is symmetric, we have $\bm{R}_{ij} \bm{R}_{ij} : \nabla \bm{u}(\bm{r}) = \bm{R}_{ij} \bm{R}_{ij} : [\nabla \bm{u}(\bm{r})]^{\mathrm{T}}$.
Therefore we can rewrite eq~\eqref{potential_energy_difference_target} as
\begin{equation}
 \label{potential_energy_difference_modified3}
  \mathcal{U}(\lbrace \bm{R}_{i}' \rbrace)
  - \mathcal{U}(\lbrace \bm{R}_{i} \rbrace)
   = \int d\bm{r} \sum_{i > j} \frac{\partial \phi_{ij}(R_{ij})}
  {\partial R_{ij}} \frac{\bm{R}_{ij} \bm{R}_{ij} }{R_{ij}} \int_{0}^{1} d \lambda \,
  \delta(\bm{r} - \bm{R}_{j} - \lambda \bm{R}_{ij}) : \bm{\epsilon}(\bm{r}).
\end{equation}
(If we do not employ the Irving-Kirkwood path, such a manipulation
cannot be done.)
By comparing eqs~\eqref{potential_energy_difference_target}
and \eqref{potential_energy_difference_modified3}, we find that 
the the potential part stress tensor field should be defined as
\begin{equation}
 \label{irving_kirkwood_stress_potential_part}
 \hat{\bm{\sigma}}^{(\mathrm{p})}(\bm{r})
  \equiv \sum_{i > j} \frac{\partial \phi_{ij}(R_{ij})}
  {\partial R_{ij}} \frac{\bm{R}_{ij} \bm{R}_{ij} }{R_{ij}} \int_{0}^{1} d \lambda \,
  \delta(\bm{r} - \bm{R}_{j} - \lambda \bm{R}_{ij}) .
\end{equation}
As expected, eq~\eqref{irving_kirkwood_stress_potential_part} is nothing but
the potential part of the Irving-Kirkwood stress tensor\cite{Irving-Kirkwood-1950,Schfield-Henderson-1982}.

The Irving-Kirkwood
stress tensor consists of  the kinetic and potential parts. The kinetic part of
the stress tensor field is localized at the particle positions\cite{Irving-Kirkwood-1950,Schfield-Henderson-1982}. In our system,
the momentum are assumed to be fully equilibrated.
The contribution of the kinetic part stress per one particle reduces to $- k_{B} T \bm{1}$\cite{Uneyama-Nakai-Masubuchi-2019}.
Thus the kinetic part
stress tensor field should be defined as
\begin{equation}
 \label{irving_kirkwood_stress_kinetic_part}
 \hat{\bm{\sigma}}^{(\mathrm{k})}(\bm{r})
  \equiv - k_{B} T \bm{1} \sum_{i} 
  \delta(\bm{r} - \bm{R}_{i}) .
\end{equation}
The total stress tensor field is then given as the sum of
eqs~\eqref{irving_kirkwood_stress_potential_part} and \eqref{irving_kirkwood_stress_kinetic_part}:
\begin{equation}
 \label{irving_kirkwood_stress}
  \hat{\bm{\sigma}}(\bm{r})
   = \hat{\bm{\sigma}}^{(\mathrm{p})}(\bm{r})
  + \hat{\bm{\sigma}}^{(\mathrm{k})}(\bm{r}).
\end{equation}
The macroscopic
stress tensor corresponds to the spatial average of eq~\eqref{irving_kirkwood_stress}.

\subsection{Linear Response Theory}

To consider the scale-dependent linear viscoelasticity, we consider the Langevin
equation with the external flow field. We assume that the dynamics
is described by the following overdamped Langevin equation:
\begin{equation}
 \label{langevin_equation}
\begin{split}
 \frac{d\bm{R}_{i}(t)}{dt}
 & = \bm{v}(\bm{R}_{i}(t),t) - \sum_{j (\neq i)} \bm{L}_{ij}(\lbrace \bm{R}_{i} \rbrace)
  \cdot \frac{\partial \mathcal{U}(\lbrace \bm{R}_{i}  \rbrace)}{\partial \bm{R}_{j}} + k_{B} T\sum_{j} \frac{\partial}{\partial \bm{R}_{j}} \cdot \bm{L}_{ij}(\lbrace \bm{R}_{i} \rbrace) \\
 & \qquad  + \sqrt{2 k_{B} T} \bm{B}_{ij}(\lbrace \bm{R}_{i}  \rbrace) \cdot \bm{w}_{j}(t),
\end{split}
\end{equation}
where $\bm{v}(\bm{r},t)$ is the flow field,
$\bm{L}_{ij}(\lbrace \bm{R}_{i}  \rbrace)$ is the mobility tensor,
$\bm{B}_{ij}(\lbrace \bm{R}_{i}  \rbrace)$ is the noise coefficient tensor,
and $\bm{w}_{i}(t)$ is the Gaussian white noise.
We interpret the Gaussian white noise in eq~\eqref{langevin_equation} in the
Ito manner\cite{Gardiner-book}.
The fluctuation-dissipation relation requires
that the following relations hold:
\begin{equation}
 \bm{B}_{ij}(\lbrace \bm{R}_{i}  \rbrace) \cdot \bm{B}_{kj}^{\mathrm{T}}(\lbrace \bm{R}_{i}  \rbrace)
  = \bm{L}_{ik}(\lbrace \bm{R}_{i}  \rbrace),
\end{equation}
\begin{equation}
 \langle \bm{w}_{i}(t) \rangle = 0, \qquad
 \langle \bm{w}_{i}(t) \bm{w}_{j}(t') \rangle = \bm{1} \delta(t - t').
\end{equation}

The dynamic equation for the time-dependent probability distribution
would be convenient when we consider the linear response.
We describe the time-dependent probability distribution for particle positions
as $P(\lbrace \bm{R}_{i} \rbrace,t)$.
From the Langevin equation \eqref{langevin_equation}, $P(\lbrace \bm{R}_{i} \rbrace,t)$
obeys the following Fokker-Planck equation\cite{Gardiner-book}:
\begin{equation}
 \label{fokker_planck_equation}
  \frac{\partial P(\lbrace \bm{R}_{i} \rbrace,t)}{\partial t}
  = \mathcal{L}_{0} P(\lbrace \bm{R}_{i} \rbrace,t)
  - \sum_{i} \frac{\partial}{\partial \bm{R}_{i}}
  \cdot [\bm{v}(\bm{R}_{i},t) P(\lbrace \bm{R}_{i} \rbrace,t)],
\end{equation}
where we have defined the equilibrium Fokker-Planck operator as
\begin{equation}
 \label{equilibrium_fokker_planck_operator}
  \mathcal{L}_{0} P(\lbrace \bm{R}_{i} \rbrace)
   \equiv \sum_{i,j} \frac{\partial}{\partial \bm{R}_{i}} \cdot 
   \left[ \bm{L}_{ij}(\lbrace \bm{R}_{i}  \rbrace) 
  \cdot \left[ \frac{\partial \mathcal{U}(\lbrace \bm{R}_{i}
  \rbrace)}{\partial \bm{R}_{j}} P(\lbrace \bm{R}_{i} \rbrace)
	+ k_{B} T \frac{\partial P(\lbrace \bm{R}_{i}
 \rbrace)}{\partial \bm{R}_{j}} \right] \right].
\end{equation}
In absence of the flow field, $\bm{v}(\bm{r},t) = 0$, the stationary
solution of the Fokker-Planck equation becomes the equilibrium distribution
\begin{equation}
 \label{equilibrium_probability_distribution}
 P_{\text{eq}}(\lbrace \bm{R}_{i} \rbrace) \equiv \frac{1}{\mathcal{Z}}
  \exp\left[ - \frac{\mathcal{U}(\lbrace \bm{R}_{i}
  \rbrace)}{k_{B} T} \right],
\end{equation}
with the partition function defined as
\begin{equation}
  \mathcal{Z} \equiv \int d\lbrace \bm{R}_{i} \rbrace \,  \exp\left[ - \frac{\mathcal{U}(\lbrace \bm{R}_{i}
  \rbrace)}{k_{B} T} \right].
\end{equation}
It is straightforward to show $\mathcal{L}_{0} P_{\text{eq}}(\lbrace \bm{R}_{i} \rbrace) = 0$.
We define the equilibrium statistical average by using eq~\eqref{equilibrium_probability_distribution} as
$\langle \dotsb \rangle_{\text{eq}} \equiv \int d\lbrace\bm{R}_{i}\rbrace \, \dotsb P_{\text{eq}}(\lbrace \bm{R}_{i} \rbrace)$.
In equilibrium, from the symmetry, the average stress tensor should be
isotropic and homogeneous. Thus the average stress tensor
can be characterized only by the pressure $p_{\text{eq}}$:
\begin{equation}
 \langle \hat{\bm{\sigma}}(\bm{r}) \rangle_{\text{eq}} = - p_{\text{eq}} \bm{1}.
\end{equation}

If a weak time-dependent flow field is applied, the distribution function is slightly
deviated from the equilibrium distribution.
Following the standard procedure\cite{Evans-Morris-book}, we express the distribution function
as the sum of the equilibrium part and the time-dependent perturbation part:
\begin{equation}
 P(\lbrace \bm{R}_{i} \rbrace,t) =
   P_{\text{eq}}(\lbrace \bm{R}_{i} \rbrace)
   +  \Delta P(\lbrace \bm{R}_{i} \rbrace,t).
\end{equation}
We assume that the flow
field is sufficiently small, and thus the perturbation part is also sufficiently small.
Then, the higher order perturbtion terms than the second order can be neglected.
We have the following equation for the pertrubation part:
\begin{equation}
 \label{fokker_planck_equation_perturbation}
  \frac{\partial \Delta P(\lbrace \bm{R}_{i} \rbrace,t)}{\partial t} =
  \mathcal{L}_{0} \Delta P(\lbrace \bm{R}_{i} \rbrace,t)
  - \sum_{i} \frac{\partial}{\partial \bm{R}_{i}}
  \cdot [\bm{v}(\bm{R}_{i},t) P_{\text{eq}}(\lbrace \bm{R}_{i} \rbrace)].
\end{equation}
Eq~\eqref{fokker_planck_equation_perturbation} can be formally solved as follows:
\begin{equation}
 \label{distribution_function_perturbation_part}
\begin{split}
 \Delta P(\lbrace \bm{R}_{i} \rbrace,t)
 & =
 - \int_{-\infty}^{t} dt' \, e^{(t - t') \mathcal{L}_{0}} 
 \sum_{i} \frac{\partial}{\partial \bm{R}_{i}} \cdot
 \left[ \bm{v}(\bm{R}_{i},t')  P_{\text{eq}}(\lbrace \bm{R}_{i} \rbrace) \right]
  \\
 & =
 - \int_{-\infty}^{t} dt' \, e^{(t - t') \mathcal{L}_{0}} 
 \bigg[ 
 \sum_{i} \left[ \frac{\partial}{\partial \bm{R}_{i}} \cdot \bm{v}(\bm{R}_{i},t') \right]
  P_{\text{eq}}(\lbrace \bm{R}_{i} \rbrace) \\
 & \qquad - \frac{1}{k_{B} T} \sum_{i} \frac{\partial \mathcal{U}(\lbrace \bm{R}_{i} \rbrace)}{\partial \bm{R}_{i}}
 \cdot
 \bm{v}(\bm{R}_{i},t') P_{\text{eq}}(\lbrace \bm{R}_{i} \rbrace)
 \bigg].
\end{split}
\end{equation}
The first term in the last line of eq~\eqref{distribution_function_perturbation_part} can be rewritten as follows:
\begin{equation}
 \label{distribution_function_perturbation_part_first_term}
 \begin{split}
  & \sum_{i} \left[ \frac{\partial}{\partial \bm{R}_{i}} \cdot \bm{v}(\bm{R}_{i},t') \right]
  P_{\text{eq}}(\lbrace \bm{R}_{i} \rbrace) \\
  & = \frac{1}{k_{B} T} \left[ \int d\bm{r}' \, k_{B} T \sum_{i} \delta(\bm{r}' -
  \bm{R}_{i}) \bm{1} : \nabla' \bm{v}(\bm{r}',t') \right]
  P_{\text{eq}}(\lbrace \bm{R}_{i} \rbrace)  \\
  & = - \frac{1}{k_{B} T} \left[ \int d\bm{r}' \, \hat{\bm{\sigma}}^{(\mathrm{k})}(\bm{r}') : \bm{\kappa}(\bm{r}',t') \right]
  P_{\text{eq}}(\lbrace \bm{R}_{i} \rbrace) ,
 \end{split}
\end{equation}
where $\nabla' \equiv \partial / \partial \bm{r}'$, and $\bm{\kappa}(\bm{r},t) \equiv [\nabla \bm{v}(\bm{r},t)]^{\mathrm{T}}$ can be
interpreted as the velocity gradient tensor field.
The second term in the last line of eq~\eqref{distribution_function_perturbation_part}
has a similar form to eq~\eqref{potential_energy_difference}. Actually, it can
be rewritten in terms of the potential part stress tensor field and the velocity gradient field:
\begin{equation}
 \label{distribution_function_perturbation_part_second_term}
 \begin{split}
  & - \frac{1}{k_{B} T} \sum_{i} \frac{\partial \mathcal{U}(\lbrace \bm{R}_{i} \rbrace)}{\partial \bm{R}_{i}}
 \cdot
 \bm{v}(\bm{R}_{i},t') P_{\text{eq}}(\lbrace \bm{R}_{i} \rbrace) \\
  & = - \frac{1}{k_{B} T} \left[ \int d\bm{r}' \, \hat{\bm{\sigma}}^{(\mathrm{p})}(\bm{r}') : \bm{\kappa}(\bm{r}',t') \right]
  P_{\text{eq}}(\lbrace \bm{R}_{i} \rbrace) .
 \end{split}
\end{equation}
By substituting eqs~\eqref{distribution_function_perturbation_part_first_term}
and \eqref{distribution_function_perturbation_part_second_term} into
eq~\eqref{distribution_function_perturbation_part},
finally the perturbation part of the distribution function can be rewritten by using the Irving-Kirkwood stress tensor field:
\begin{equation}
 \label{distribution_function_perturbation_part_final}
 \Delta P(\lbrace \bm{R}_{i} \rbrace,t)
 = \frac{1}{k_{B} T}  \int d\bm{r}' \int_{-\infty}^{t} dt' \, e^{(t - t') \mathcal{L}_{0}} 
 \left[ \hat{\bm{\sigma}}(\bm{r}') : \bm{\kappa}(\bm{r}',t')
 P_{\text{eq}}(\lbrace \bm{R}_{i} \rbrace) \right].
\end{equation}
When we monitor the average stress tensor field under flow, we will observe a time-dependent
tensor field $\bm{\sigma}(\bm{r},t)$. From eq~\eqref{distribution_function_perturbation_part_final}, it can be
expressed as
\begin{equation}
 \label{average_stress_tensor_under_flow}
  \begin{split}
   & \bm{\sigma}(\bm{r},t) + p_{\text{eq}} \bm{1} = \int d\lbrace \bm{R}_{i} \rbrace \, \hat{\bm{\sigma}}(\bm{r}) 
\Delta P(\lbrace \bm{R}_{i} \rbrace,t) \\
   & =
    \frac{1}{k_{B} T} \int d\bm{r}' \int_{-\infty}^{t} dt' \int d\lbrace \bm{R}_{i} \rbrace \, 
   [e^{(t - t') \mathcal{L}^{\dagger}_{0}} \hat{\bm{\sigma}}(\bm{r})] \left[ 
   \hat{\bm{\sigma}}(\bm{r}') :
\bm{\kappa}(\bm{r}',t') P_{\text{eq}}(\lbrace \bm{R}_{i} \rbrace)
   \right] \\
   & = \frac{1}{k_{B} T} \int d\bm{r}' \int_{-\infty}^{t} dt' \langle
   \hat{\bm{\sigma}}(\bm{r}, t - t')  \hat{\bm{\sigma}}(\bm{r}')
 \rangle_{\text{eq}} : \bm{\kappa}(\bm{r}',t').
  \end{split}
\end{equation}
Here,
$\mathcal{L}^{\dagger}_{0}$ is the adjoint Fokker-Planck
operator, and $\hat{\bm{\sigma}}(\bm{r},t) \equiv e^{t \mathcal{L}_{0}^{\dagger}} \hat{\bm{\sigma}}(\bm{r})$
is the time-shifted stress tensor field.
From eq~\eqref{average_stress_tensor_under_flow}, we find that
the response of the stress tensor field to the applied velocity gradient
field $\bm{\kappa}(\bm{r},t)$ is characterized by
the following relaxation modulus tensor:
\begin{equation}
 \label{relaxation_modulus_tensor}
 \bm{\Lambda}(\bm{r},t) \equiv \frac{1}{k_{B} T} \langle
   \hat{\bm{\sigma}}(\bm{r}, t)  \hat{\bm{\sigma}}(0,0)
 \rangle_{\text{eq}}.
\end{equation}
Eq~\eqref{relaxation_modulus_tensor} can be interpreted as the distance-dependent relaxation
modulus. $\bm{\Lambda}(\bm{r},t)$ represents the response of the stress tensor field
at position $\bm{r}$ and time $t$, to the perturbation of the velocity gradient tensor field
at position $0$ and time $0$.

The (spatial) Fourier transform of eq~\eqref{relaxation_modulus_tensor}
would be convenient. We introduce the
Fourier-transformed relaxation modulus:
\begin{equation}
 \label{relaxation_modulus_tensor_fourier_transform}
 \bm{\Lambda}(\bm{k},t)  \equiv 
  \int d\bm{r} \, e^{-i \bm{r} \cdot\bm{k}} 
   \bm{\Lambda}(\bm{r},t).
\end{equation}
Eq~\eqref{relaxation_modulus_tensor_fourier_transform}
can be interpreted as the wavenumber-dependent relaxation modulus.
The Fourier transform of eq~\eqref{average_stress_tensor_under_flow} is
\begin{equation}
 \bm{\sigma}(\bm{k},t) + p_{\text{eq}} \delta(\bm{k}) \bm{1}
  = \int_{-\infty}^{t} dt' \, \bm{\Lambda}(\bm{k},t - t') : \bm{\kappa}(\bm{k},t'),
\end{equation}
where $\bm{\sigma}(\bm{k},t)$ and $\bm{\kappa}(\bm{k},t)$ are the
Fourier transforms of $\bm{\sigma}(\bm{r},t)$ and $\bm{\kappa}(\bm{r},t)$.

Eq~\eqref{relaxation_modulus_tensor} has almost the same form as the
Green-Kubo formula for macroscopic relaxation modulus. Thus we
interpret eq~\eqref{relaxation_modulus_tensor} as the scale-dependent
version of the Green-Kubo formula. Here it should be pointed that Evans\cite{Evans-1981}
derived a similar but different linear response formula. Evans derived
the linear response based on the generalized hydrodynamics and claimed that
the correlation function of the the transverse momentum current field
gives the relaxation modulus. Thus the dynamics of individual particles is
not explicitly considered. In contrast, our derivation is based on
the Langevin equation. Our system does not have
the degrees of freedom of momenta, and thus we cannot directly
apply Evans's approach to our system.
For particle-based systems, our approach seems to
be physically natural. 

\subsection{Transverse and Longitudinal Modes}

\begin{figure}[tb]
 \centering
 \includegraphics[width=0.3\linewidth,clip]{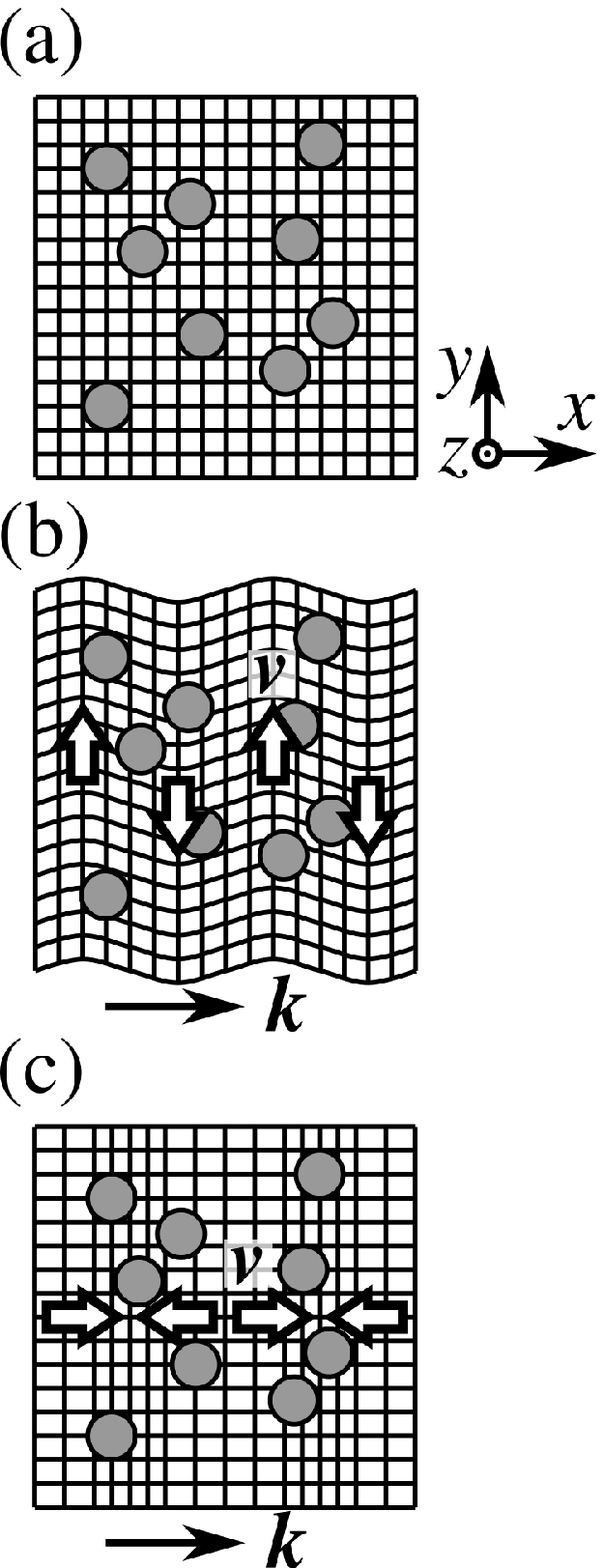}
\caption{\label{deformation_with_finite_wavenumber}
 The deformation modes with a finite wavenumber.
 Gray circles represent particles. Grids are shown as the guide for eye.
 (a) The reference state without any deformations. The $x-$, $y$-, and $z$-directions are
 shown with arrows.
 (b) The transverse deformation mode. The wavenumber vector $\bm{k}$
 is parallel to the $x$-axis whereas the velocity vector $\bm{v}$ is parallel to the $y$-axis.
 $\bm{k}$ and $\bm{v}$ are orthogonal.
 (c) The longitudinal deformation mode. Both the wavenumber vector $\bm{k}$
 and the velocity vector $\bm{v}$ are parallel to the $x$-axis.}
\end{figure}

At the linear response regime, a deformation field can be decomposed into
deformation fields with different wavenumber vectors.
Here we consider a deformation field with a single wavenumber vector $\bm{k}$.
Without loss of generality,
we can set the wavenumber vector $\bm{k} = k \bm{e}_{x}$ with $\bm{e}_{x}$
with being the unit vector in the $x$-direction, to analyze the relaxation
modulus. The wavenumber-dependent fields such as $\bm{\sigma}(\bm{k},t)$
can be interpreted as a function of $k$ such as $\bm{\sigma}(k,t)$.
From the symmetry, we need to consider only two types of
velocity gradient fields: the transverse mode and the longitudinal mode.
Figure~\ref{deformation_with_finite_wavenumber} illustrates the transverse
and longitudinal deformation modes.
The velocity fields of the transverse and longitudinal modes are given as,
for example,
\begin{align}
  \label{velocity_fields_transverse}
 \bm{v}^{(\mathrm{t})}(\bm{r},t) & = v(t) \sin (k r_{x}) \bm{e}_{y}, \\
 \label{velocity_fields_longitudinal}
  \bm{v}^{(\mathrm{l})}(\bm{r},t) & = v(t) \sin (k r_{x}) \bm{e}_{x},
\end{align}
where the superscripts ``(t)'' and ``(l)'' represent the transverse and
longitudinal modes, respectively.
$v(t)$ is the time-dependent velocity amplitude, and $\bm{e}_{y}$ is the unit vector in the $y$-direction.
The velocity gradient tensor fields which corresponds to eqs~\eqref{velocity_fields_transverse}
and \eqref{velocity_fields_longitudinal} are
\begin{align}
  \label{velocity_gradient_fields_transverse}
  \bm{\kappa}^{(\mathrm{t})}(\bm{r},t) & = 
  \kappa(t) \cos (k r_{x}) \bm{e}_{y} \bm{e}_{x}, \\
 \label{velocity_gradient_fields_longitudinal}
    \bm{\kappa}^{(\mathrm{l})}(\bm{r},t) & = 
\kappa(t) \cos (k r_{x}) \bm{e}_{x} \bm{e}_{x},
\end{align}
where $\kappa(t) \equiv k v(t)$ can be interpreted as the amplitude
of the velocity gradient.

We have only two independent components for
the relaxation modulus tensor for an isotropic material\cite{Landau-Lifshitz-book}.
For the transverse mode, the response of the shear stress field would be 
useful. From eq~\eqref{relaxation_modulus_tensor_fourier_transform} we have
$\sigma_{yx}(k,t) = \Lambda_{yxyx}(k,t) \kappa(t) \cos(k r_{x})$.
We define the wavenumber-dependent shear relaxation modulus as
\begin{equation}
 \label{shear_modulus_definition}
 G(k,t) \equiv  \Lambda_{yxyx}(k,t).
\end{equation}
For the longitudinal mode, the response of the normal stress fields would be
useful. From eq~\eqref{relaxation_modulus_tensor_fourier_transform}, 
$\sigma_{xx}(k,t) =  \Lambda_{xxxx}(k,t) \kappa \cos(k r_{x})$.
$\Lambda_{xxxx}(k,t)$ can be interpreted as the wavenumber-dependent
longitudinal modulus.
Then we can define the wavenumber-dependent bulk relaxation modulus as
\begin{equation}
 \label{bulk_modulus_definition}
 K(k,t) \equiv \Lambda_{xxxx}(k,t) - \frac{4}{3} G(k,t),
\end{equation}
with $G(k,t)$ given by eq~\eqref{shear_modulus_definition}.
Alternatively, we can utilize the sum of all the normal components to calculate
the bulk modulus:
\begin{equation}
 \label{bulk_modulus_definition_modified}
 K(k,t) = \frac{1}{3} \sum_{\alpha = x,y,z} \Lambda_{\alpha\alpha xx}(k,t) .
\end{equation}
The storage and loss moduli are defined as the (temporal) Fourier transforms
of eqs~\eqref{shear_modulus_definition} and \eqref{bulk_modulus_definition}:
\begin{align}
 \label{shear_complex_moduli_definition}
 G'(k,\omega) + i G''(k,\omega) \equiv i \omega \int_{0}^{\infty} dt \, e^{-i \omega t} G(k,t), \\
 \label{bulk_complex_moduli_definition}
 K'(k,\omega) + i K''(k,\omega) \equiv i \omega \int_{0}^{\infty} dt \, e^{-i \omega t} K(k,t).
\end{align}
Eqs~\eqref{shear_complex_moduli_definition} and \eqref{bulk_complex_moduli_definition}
describe the responses of the wavenumber-dependent stress to the temporarily oscillating wavenumber-dependent velocity fields.
The macroscopic moduli are recovered at $k = 0$.
For example, the macroscopic shear relaxation modulus is given as
$G(t) = G(0,t)$.

\section{EXAMPLES}

\subsection{Non-Interacting Brownian Particles}
\label{non_interacting_brownian_particles}

As a simple example, we consider non-interacting Brownian particles.
We express the number
of Brownian particles as $M$ and the system size as $V$. The
number density of particles is $\nu = M / V$.

There is no interaction potential in this system and thus we have
only the kinetic part for the stress tensor field. Also, due to
the non-interacting nature, the contributions of individual
particles are statistically independent. Then what we need to
consider is the kinetic part stress tensor for just a single particle:
\begin{equation}
 \label{stress_tensor_brownian_particle}
 \hat{\bm{\sigma}}(\bm{r},t)
 = \hat{\bm{\sigma}}^{(\mathrm{k})}(\bm{r},t)
  = - k_{B} T \bm{1} \delta(\bm{r} - \bm{R}(t)),
\end{equation}
where $\bm{R}$ is the position of the Brownian particle.

The dynamic equation for the particle position $\bm{R}$ is
\begin{equation}
 \label{langevin_equation_brownian_particle}
 \frac{d\bm{R}(t)}{dt} = \sqrt{\frac{2k_{B}T}{\zeta}}\bm{w}(t),
\end{equation}
where $\zeta$ is the friction coefficient and $\bm{w}(t)$ is the Gaussian
white noise.
The propagator for the position is calculated to be
\begin{equation}
 \label{propagator_brownian_particle}
 \begin{split}
  W(\bm{r},t|\bm{r}',0)
  & = \left\langle \delta(\bm{r} - \bm{R}(t)) \delta(\bm{r}' - \bm{R}(0)) \right\rangle \\
  & = \left(\frac{1}{4 \pi D t}\right)^{3/2}
 \exp\left[ - \frac{(\bm{r} - \bm{r}')^{2}}{4 D t} \right],
 \end{split}
\end{equation}
where $D \equiv k_{B} T / \zeta$ is the diffusion coefficient.
The equilibrium distribution for $\bm{R}$ is trivial: $P_{\text{eq}}(\bm{R}) = 1 / V$.

The relaxation modulus of a single particle can be calculated
by using eqs~\eqref{propagator_brownian_particle} and \eqref{stress_tensor_brownian_particle}.
\begin{equation}
 \label{relaxation_modulus_brownian_particle_single}
 \begin{split}
   \bm{\Lambda}^{(\text{single})}(\bm{r},t) 
  &  = \frac{1}{k_{B} T}
  \langle \hat{\bm{\sigma}}(\bm{r},t) \hat{\bm{\sigma}}(0,0) \rangle_{\text{eq}} \\
  & = \frac{k_{B} T}{V} \bm{1} \bm{1} \int d\bm{R}d\bm{R}' \,
   W(\bm{R},t|\bm{R}',0) P_{\text{eq}}(\bm{R}') \delta(\bm{r} - \bm{R}) \delta(\bm{R}') \\
  & = \frac{k_{B} T}{V} \bm{1} \bm{1}
 \left(\frac{1}{4 \pi D t}\right)^{3/2}
\exp\left( - \frac{\bm{r}^{2}}{4 D t} \right).
 \end{split}
\end{equation}
By collecting the contributions of $M$ particles, the relaxation modulus
of the non-interacting Brownian particles becomes
\begin{equation}
 \label{relaxation_modulus_brownian_particle}
   \bm{\Lambda}(\bm{r},t) 
  = M \bm{\Lambda}^{(\text{single})}(\bm{r},t)  = \nu k_{B} T \bm{1} \bm{1}
 \left(\frac{1}{4 \pi D t}\right)^{3/2}
\exp\left( - \frac{\bm{r}^{2}}{4 D t} \right).
\end{equation}
Then the Fourier transform of eq~\eqref{relaxation_modulus_brownian_particle}
with $\bm{k} = k \bm{e}_{x}$ is
\begin{equation}
 \label{relaxation_modulus_brownian_particle_fourier_transform}
   \bm{\Lambda}(k,t) 
   = \int d\bm{r} e^{-i k r_{x}}   \bm{\Lambda}(\bm{r},t) 
   = \nu k_{B} T \bm{1} \bm{1}
\exp(  - D k^{2} t) .
\end{equation}
From eq~\eqref{relaxation_modulus_brownian_particle_fourier_transform}, the shear relaxation modulus is zero: $G(k,t) = 0$. The bulk relaxation modulus
is
\begin{equation}
 \label{bulk_relaxation_modulus_brownian_particle}
 K(k,t) = \nu k_{B} T \exp(  - D k^{2} t) .
\end{equation}
The (temporal) Fourier-transform of eq~\eqref{bulk_relaxation_modulus_brownian_particle} gives
the wavenumber-dependent bulk storage and loss moduli:
\begin{equation}
 K'(k,\omega) = \nu k_{B} T\frac{(\omega / D k^{2})^{2}}{1 + (\omega / D k^{2})^{2}}, \qquad
 K''(k,\omega) = \nu k_{B} T\frac{\omega / D k^{2}}{1 + (\omega / D k^{2})^{2}}.
\end{equation}

From eq~\eqref{bulk_relaxation_modulus_brownian_particle}, we find that the bulk relaxation modulus at a finite wavenumber
decays exponentially. The relaxation time becomes a function of the wavenumber $k$,
as $\tau(k) \equiv 1 / D k^{2}$.
This relation is physically reasonable
because the relaxation at the finite wavenumber should be governed by the
diffusion of Brownian particles, and the large wavenumber modes generally
decay faster than the low wavenumber modes in the diffusion dynamics.
By setting $k = 0$, we have the macroscopic bulk relaxation modulus:
$K(t) = K(0,t) = \nu k_{B} T$. This simply means that the bulk modulus is constant and
does not relax. Macroscopically, non-interacting Brownian particles simply
behaves as an ideal gas. There is essentially no macroscopic relaxation.
However, at the finite wavenumber, they exhibit the relaxation by the diffusion.

One may consider that the wavenumber dependence of eq~\eqref{bulk_relaxation_modulus_brownian_particle}
is similar to that of a dynamic structure factor in scattering experiments\cite{Ewen-Richter-1997}.
This is rather natural
because both the dynamic structure factor and the bulk relaxation modulus
reflect the diffusion dynamics.
The kinetic part of the stress tensor field is localized at the particle positions,
and the relaxation modulus becomes the two-point correlation function for the
particle position, as observed in eq~\eqref{relaxation_modulus_brownian_particle_single}.


\subsection{Harmonic Dumbbell Model}
\label{harmonic_dumbbell}

As another simple example, we consider the harmonic dumbbell model. We
consider a system which consists of non-interacting harmonic dumbbells\cite{Ottinger-book,Kroger-2004}.
We express the system size as $V$ and assume that $M$ dumbbells are
uniformly dispersed in the system. The number density of dumbbells
is $\nu = M / V$.

Due to the non-interacting
nature, the information on the dynamics of a single dumbbell is sufficient
for us to calculate the linear response, in the same way as
Sec.~\ref{non_interacting_brownian_particles}.
We express the positions of two particles which consist a dumbbell as
$\bm{R}_{1}$ and $\bm{R}_{2}$. The interaction potential is
$\phi(\bm{R}_{1} - \bm{R}_{2}) = (3 k_{B} T / 2 b^{2}) (\bm{R}_{1} - \bm{R}_{2})^{2}$
with $b$ being the average bond size.
We employ the overdamped Langevin
equation for the time-evolution:
\begin{equation}
 \label{langevin_equation_dumbbell}
 \frac{d\bm{R}_{i}(t)}{dt} = - \frac{1}{\zeta} 
  \frac{\partial \phi(\bm{R}_{1}(t) - \bm{R}_{2}(t))}{\partial \bm{R}_{i}(t)} + \sqrt{\frac{2 k_{B}T}{\zeta}}\bm{w}_{i}(t),
\end{equation}
where $\zeta$ is the friction coefficient,
and $\bm{w}_{i}(t)$ is the Gaussian white noise.
The noise $\bm{w}_{i}(t)$ satisfies the fluctuation-dissipation relation:
\begin{equation}
 \langle \bm{w}_{i}(t) \rangle = 0, \qquad
 \langle \bm{w}_{i}(t) \bm{w}_{j}(t') \rangle =  \delta_{ij} \bm{1} \delta(t - t').
\end{equation}

The dynamics of the harmonic dumbbell model can be solved analytically. We
introduce the center of mass position and the bond vector
$ \bm{R}(t) \equiv [\bm{R}_{1}(t) + \bm{R}_{2}(t)] / 2$ and
$\bm{Q}(t) \equiv \bm{R}_{1}(t) - \bm{R}_{2}(t)$.
Then, we can rewrite the Langevin equation \eqref{langevin_equation_dumbbell}
into the Langevin equations for the center of mass position and the bond vector:
\begin{align}
 \label{langevin_equation_cm}
 \frac{d \bm{R}(t)}{dt} & = \sqrt{\frac{k_{B} T}{\zeta}} \bm{w}_{+}(t), \\
 \label{langevin_equation_bond}
 \frac{d \bm{Q}(t)}{dt} & = - \frac{6 k_{B} T}{\zeta b^{2}} \bm{Q}(t) + 
 \sqrt{\frac{4 k_{B}T}{\zeta}} \bm{w}_{-}(t),
\end{align}
with $\bm{w}_{\pm}(t) \equiv [\bm{w}_{1}(t) \pm \bm{w}_{2}(t)] / \sqrt{2}$.

We need the explicit expression of the stress correlation function to
calculate the relaxation modulus. It can be calculated by using the
equilibrium distribution function and the propagators.
The equilibrium distribution is simply given as
\begin{equation}
 \label{equilibrium_distribution}
 P_{\text{eq}}(\bm{R},\bm{Q}) = \frac{1}{V}
 \left( \frac{3}{2 \pi b^{2}} \right)^{3/2}
 \exp\left( - \frac{3 \bm{Q}^{2}}{2 b^{2}} \right).
\end{equation}
The propagators for the center of mass and the bond are statistically
independent.
Eq~\eqref{langevin_equation_bond} describes the Ornstein-Uhlenbeck process\cite{vanKampen-book}.
The solution is
\begin{equation}
 \bm{Q}(t) = e^{-t / \tau} \bm{Q}(0) + \sqrt{\frac{4 k_{B} T}{\zeta}}\int_{0}^{t} dt' \, e^{-(t - t')
  / \tau} \bm{w}_{-}(t'),
\end{equation}
where $\tau \equiv {\zeta b^{2}}/{6 k_{B} T}$ is the relaxation time.
The propagator for $\bm{Q}$ is given as\cite{vanKampen-book}
\begin{equation}
 \label{propagator_bond}
  \begin{split}
  W_{Q}(\bm{q},t|\bm{q}',0)
  & = \langle \delta(\bm{q} - \bm{Q}(t)) \delta(\bm{q}' - \bm{Q}(0)) \rangle \\
   & = \left[
     \frac{3}{2 \pi b^{2} (1 - e^{-2 t / \tau})}
    \right]^{3/2}
  \exp
  \left[ - \frac{3 ( \bm{q} - e^{-t / \tau} \bm{q}')^{2}}{2 b^{2} (1 - e^{-2 t / \tau})}
  \right].
\end{split}
\end{equation}
The propagator for $\bm{R}$ is essentially the same as that calculated in
Sec.~\ref{non_interacting_brownian_particles}.
By replacing $D$ in \eqref{propagator_brownian_particle}
by $k_{B} T / 2 \zeta = b^{2}  / 12 \tau$,
the propagator for $\bm{R}$ becomes
\begin{equation}
 \label{propagator_cm}
  W_{R}({\bm{r}},t|{\bm{r}}',0)
  = \left( \frac{3 \tau}{\pi  b^{2} t}\right)^{3/2}
  \exp\left[ - \frac{3 \tau ({\bm{r}} - {\bm{r}}')^{2}}{b^{2} t} \right]. 
\end{equation}

We calculate the stress correlation functions by using
eqs~\eqref{equilibrium_distribution}, \eqref{propagator_bond},
and \eqref{propagator_cm}.
The kinetic and potential parts of the stress tensor field for a single dumbbell are
\begin{align}
 \hat{\bm{\sigma}}^{(\mathrm{k})}(\bm{r},t)
 & = - k_{B} T \bm{1} \sum_{\mu = \pm 1/2} \delta(\bm{r} - \bm{R}(t) - \mu \bm{Q}(t)), \\
  \hat{\bm{\sigma}}^{(\mathrm{p})}(\bm{r},t)
  & = \frac{3 k_{B} T}{b^{2}} \bm{Q}(t) \bm{Q}(t)
  \int_{-1/2}^{1/2} d \mu \,
  \delta(\bm{r} - {\bm{R}}(t) - \mu \bm{Q}(t)) .
\end{align}
The kinetic part does not have the shear component.
We can ignore
the kinetic part when we calculate the shear relaxation modulus.
For the bulk relaxation modulus,
however, we need the
contribution of the kinetic part stress tensor. This makes the calculations for
the bulk relaxation modulus very lengthy and complicated.
In this work, therefore, we limit ourselves to the shear relaxation modulus and do not
go into detail about the bulk relaxation modulus.

\begin{figure}[tb]
 \centering
 \includegraphics[width=.7\linewidth,clip]{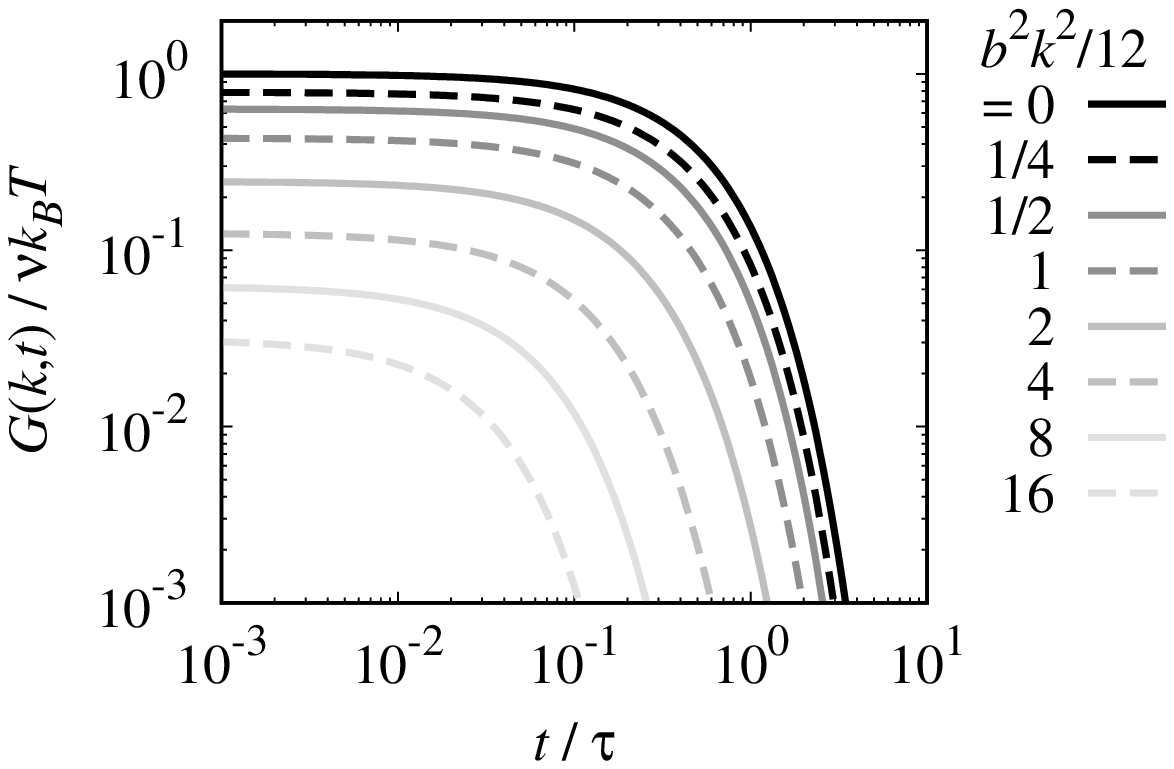}
\caption{
 The wavenumber-dependent shear relaxation modulus $G(k,t)$ for non-interacting harmonic dumbbells.
 $G(k,t)$ with several different values of $k$ are shown as functions of $t$.
 $G(0,t)$ corresponds to the macroscopic shear relaxation modulus $G(t) = \nu k_{B} T e^{- 2 t / \tau}$.
 \label{shear_relaxation_modulus_dumbbell_plot}}
\end{figure}

To calculate the shear relaxation modulus $G(k,t)$, we first
calculate $\Lambda_{yxyx}(\bm{r},t)$:
\begin{equation}
 \label{shear_relaxation_modulus_harmonic_dumbbell}
 \begin{split}
  \Lambda_{yxyx}(\bm{r},t)
  & = \frac{M}{k_{B} T} \int d\bm{R}d\bm{R}'d\bm{Q}d\bm{Q}' \,
  W_{R}({\bm{R}},t|{\bm{R}}',0) W_{Q}(\bm{Q},t|\bm{Q}',0)
  P_{\text{eq}}(\bm{R},\bm{Q}) \\
  & \qquad \times \frac{9 k_{B}^{2} T^{2}}{b^{4}} Q_{y} Q_{x} Q_{y}' Q_{x}'
  \int_{-1/2}^{1/2} d \mu 
  \int_{-1/2}^{1/2} d \mu' \,
  \delta(\bm{r} - {\bm{R}} - \mu \bm{Q})
  \delta(- {\bm{R}}' - \mu \bm{Q}') \\
  & = \frac{9 \nu k_{B} T}{b^{4}} 
\left[
     \frac{27 \tau}{4 \pi^{3} b^{6} t (1 - e^{-2 t / \tau})}
    \right]^{3/2}   \int_{-1/2}^{1/2} d \mu  \int_{-1/2}^{1/2} d \mu' 
  \int d\bm{Q}d\bm{Q}' \, Q_{x} Q_{y} Q_{x}' Q_{y}' \\
  & \qquad \times 
  \exp\left[ 
  - \frac{3}{b^{2}}
  \left[ 
   \frac{\tau (\bm{r} - \mu \bm{Q} + \mu' {\bm{Q}}')^{2}}{t}  
  + \frac{\bm{Q}^{2} - 2 e^{-t / \tau} \bm{Q} \cdot \bm{Q}' + {\bm{Q}'}^{2}}{2 (1 - e^{-2 t / \tau})}
  \right]
  \right] .
 \end{split}
\end{equation}
The Fourier transform of eq~\eqref{shear_relaxation_modulus_harmonic_dumbbell}
with $\bm{k} = k \bm{e}_{x}$ becomes
\begin{equation}
 \label{shear_relaxation_modulus_harmonic_dumbbell_fourier_transform}
 \begin{split}
  G(k,t) & = \int d\bm{r} \, e^{-i k r_{x}}   \Lambda_{yxyx}(\bm{r},t) \\
  & = \nu k_{B} T 
  \int_{-1/2}^{1/2} d \mu  \int_{-1/2}^{1/2} d \mu' \,
  \left[ e^{-t / \tau} 
  +  (\mu  - \mu' e^{-t / \tau}) (\mu' - \mu  e^{-t / \tau} ) \frac{b^{2} k^{2}}{3}
  \right]\\
  & \qquad \times 
  \exp\left[ - \frac{t}{\tau}
  -  \left(\frac{t}{2 \tau} + \mu^{2} - 2 \mu \mu'  e^{-t / \tau}
    + {\mu'}^{2} \right) \frac{b^{2} k^{2}}{6}
  \right]. 
 \end{split}
\end{equation}
After long calculations, we have the following simple form as the
wavenumber-dependent shear relaxation modulus:
\begin{equation}
 \label{shear_relaxation_modulus_harmonic_dumbbell_fourier_transform_final}
  G(k,t) 
   =   \nu k_{B} T 
  \exp\left[-\left(1 + \frac{b^{2} k^{2}}{12}\right)\frac{t}{\tau} 
  - \frac{ b^{2} k^{2}}{12} \right]
  \frac{12}{b^{2} k^{2}} 
  \sinh \left( \frac{b^{2} k^{2}}{12}  e^{-t / \tau}  \right) .
\end{equation}
The detailed calculations for eqs~\eqref{shear_relaxation_modulus_harmonic_dumbbell_fourier_transform} and \eqref{shear_relaxation_modulus_harmonic_dumbbell_fourier_transform_final} are summarized in Appendix~\ref{detailed_calculations}.
We show the shear relaxation modulus $G(k,t)$ with several different wavenumber values in Figure~\ref{shear_relaxation_modulus_dumbbell_plot}.
If the wavenumber $k$ is sufficiently small compared with the inverse of
the dumbbell size ($b^{2}k^{2} / 12 \ll 1$), $G(k,t)$ can be well approximated
by the macroscopic shear relaxation modulus.
Actually, by taking the limit of $k \to 0$, we recover the macroscopic shear
relaxation modulus for the dumbbell model: $G(t) = G(0,t) = \nu k_{B} Te^{-2t / \tau}$.
However, if the
wavenumber is not sufficiently small, we find that $G(k,t)$ depends on the wavenumber rather strongly.
As the wavenumber increases, the shear relaxation modulus decreases and the
relaxation time becomes short. The shape of $G(k,t)$ seems to be close to the
single exponential for any wavenumbers.

The decrease of the shear relaxation modulus as increasing the wavenumber
can be shown analytically.
From eq~\eqref{shear_relaxation_modulus_harmonic_dumbbell_fourier_transform_final}, we find that the
shear relaxation modulus at the short-time limit simply becomes
\begin{equation}
 \label{shear_relaxation_modulus_harmonic_dumbbell_fourier_transform_short_time}
 G(k,0) = \nu k_{B} T \frac{6 (1 - e^{- b^{2} k^{2} / 6})}{b^{2}k^{2}} .
\end{equation}
Eq~\eqref{shear_relaxation_modulus_harmonic_dumbbell_fourier_transform_short_time} is a
monotonically decreasing function of $k$, and thus the shear modulus decreases
as the wavenumber increases.
For sufficiently large wavenumber ($b^{2}k^{2} / 12 \gg 1$)
and relatively short-time scale ($t / \tau \ll 1$),
$G(k,t)$  can be well approximated by a single exponential decay:
\begin{equation}
  G(k,t) 
   \approx  \nu k_{B} T \frac{6}{b^{2}k^{2}}
  \exp\left[-\left(1 + \frac{b^{2} k^{2}}{6}\right) \frac{t}{\tau}  \right].
\end{equation}
Thus the relaxation is accelerated as the wavenumber increases.
The wavenumber-dependent effective relaxation time is estimated as $\tau_{\text{eff}}(k) \approx \tau / (1 + b^{2} k^{2} / 6)$.
This is consistent with Figure~\ref{shear_relaxation_modulus_dumbbell_plot}.
At the long-time scale, $G(k,t)$ switches to another single exponential decay form.
But the value of $G(k,t)$ at such a long-time scale is very small and the deviation
at the long-time region is practically negligible.

\begin{figure}[tb]
 \centering
 \includegraphics[width=.7\linewidth,clip]{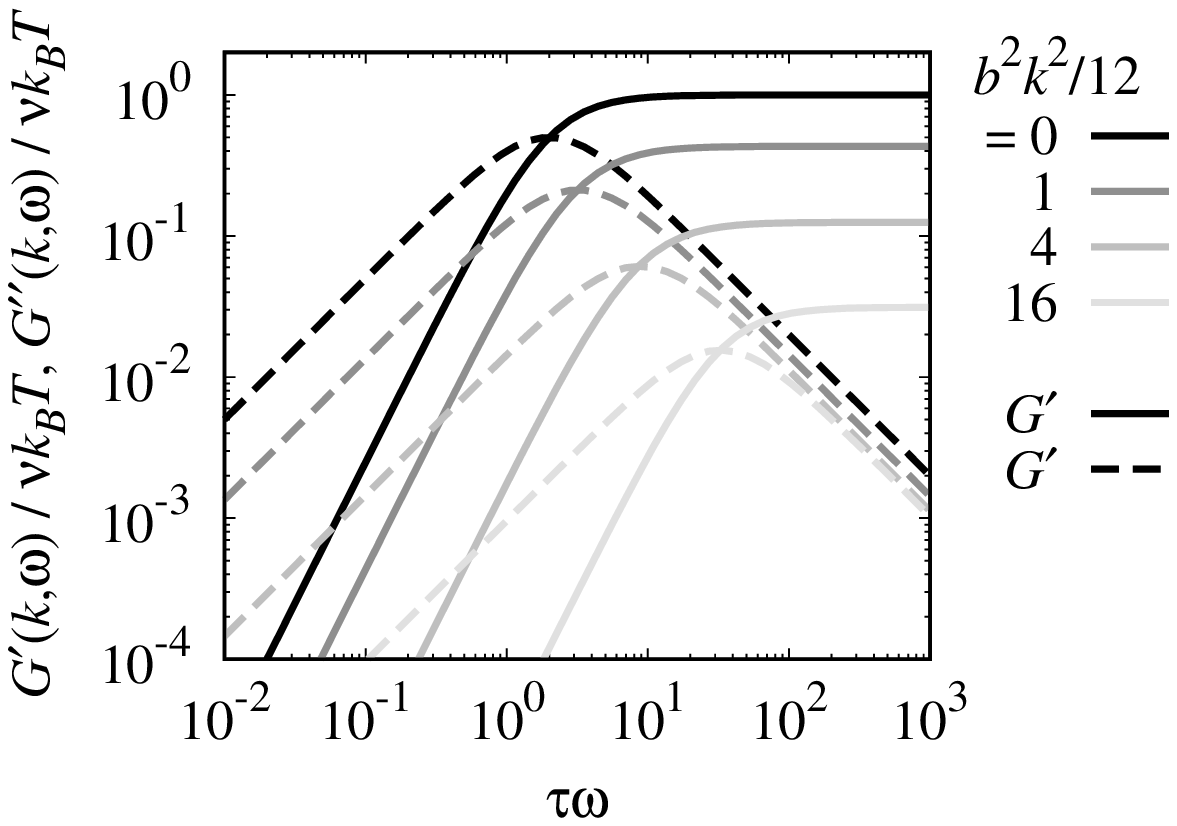}
\caption{
 The wavenumber-dependent shear storage and loss moduli $G'(k,\omega)$ and $G''(k,\omega)$ for non-interacting harmonic dumbbells.
 $G'(k,\omega)$ and $G''(k,\omega)$ are numerically calculated from $G(k,t)$.
 $G'(0,\omega)$ and $G''(0,\omega)$ correspond to the macroscopic shear storage and loss moduli, respectively.
 \label{shear_storage_and_loss_moduli_dumbbell_plot}}
\end{figure}

The wavenumber-dependent shear storage and loss moduli, $G'(k,\omega)$ and $G''(k,\omega)$,
can be calculated
by combining eqs~\eqref{shear_complex_moduli_definition} and \eqref{shear_relaxation_modulus_harmonic_dumbbell_fourier_transform_final}.
The Fourier transform of eq~\eqref{shear_relaxation_modulus_harmonic_dumbbell_fourier_transform_final} cannot be evaluated analytically, and thus
we calculate $G'(k,\omega)$ and $G''(\omega,k)$ numerically. We use the double exponential formula\cite{Ooura-Mori-1991} to accurately
calculate the integral over $\omega$ in eq~\eqref{shear_complex_moduli_definition}.
We show the shear storage and loss moduli, $G'(k,\omega)$ and $G''(k,\omega)$, in Figure~\ref{shear_storage_and_loss_moduli_dumbbell_plot}.
As expected from the $G(k,t)$ data in Figure~\ref{shear_relaxation_modulus_dumbbell_plot},
the shapes of $G'(k,\omega)$ and $G''(k,\omega)$ seem to be
similar to those of the single Maxwell model.

\section{CONCLUSIONS}

We derived the linear response theory the for scale-dependent
linear viscoelasticity of overdamped Brownian particle systems.
The dynamics of the system is assumed to be governed by the
overdamped Langevin equation, and there is no degree of freedom for
the momenta.
We showed that the Irving-Kirkwood stress tensor field can be employed
as the microscopic stress tensor field for Brownian particle systems.
Following the standard procedure,
we obtained the expression for the scale-dependent
relaxation modulus tensor. It can be interpreted as the
scale-dependent version of the Green-Kubo formula.

From the symmetry, for isotropic and homogeneous systems, there
are only two independent components for the relaxation modulus tensor field.
The transverse deformation mode gives the wavenumber-dependent
shear relaxation modulus $G(k,t)$, and the longitudinal deformation
gives the wavenumber-dependent bulk modulus $K(k,t)$. The macroscopic
relaxation modulus is recovered at $k = 0$.

As simple examples, we calculated the wavenumber-dependent
relaxation moduli for the non-interacting Brownian particles and
the harmonic dumbbell model. For the non-interacting Brownian particles,
we obtained the single exponential type bulk relaxation modulus $K(k,t)$.
The relaxation time depends on the diffusion coefficient and the
wavenumber. For the harmonic dumbbell model, we obtained the
wavenumber-dependent shear relaxation modulus $G(k,t)$. 
$G(k,t)$ has non-exponential form except $k = 0$ (eq~\eqref{shear_relaxation_modulus_harmonic_dumbbell_fourier_transform_final}),
but can be well approximated by a single exponential form.
The relaxation time and the modulus decrease as the wavenumber increases.
The fact that even these simple models exhibit nontrivial scale-dependent
linear viscoelasticity implies that more complex models and real
systems exhibit much more complex scale-dependent linear viscoelasticity.
Molecular dynamics simulations of scale-dependent relaxation moduli
for well-known systems such as polymer melts would be interesting works.
Theoretical analyses for the relation between the MSD of a colloidal
particle and the scale-dependent relaxation moduli would be also interesting.

\section*{ACKNOWLEDGMENT}

This work was supported by JST, PRESTO Grant Number JPMJPR1992, Japan,
Grant-in-Aid (KAKENHI) for Scientific Research Grant B No.~JP19H01861,
and Grant-in-Aid (KAKENHI) for Transformative Research Areas B JP20H05736.

\appendix

\section*{APPENDIX}

\section{Detailed Calculations}
\label{detailed_calculations}

In this appendix, we show the detailed calculations for the wavenumber-dependent
shear relaxation modulus $G(k,t)$ for the harmonic dumbbell model.
Although the wavenumber-dependent shear relaxation modulus $G(k,t)$ for the harmonic
dumbbell model can be calculated analytically, the calculations
are very lengthy.
We show calculations for eqs~\eqref{shear_relaxation_modulus_harmonic_dumbbell_fourier_transform}
and \eqref{shear_relaxation_modulus_harmonic_dumbbell_fourier_transform_final} in what follows.

First we show the calculations for eq~\eqref{shear_relaxation_modulus_harmonic_dumbbell_fourier_transform},
which is the
Fourier transform of eq~\eqref{shear_relaxation_modulus_harmonic_dumbbell}.
In eq~\eqref{shear_relaxation_modulus_harmonic_dumbbell},
the $\bm{r}$-dependent part is only the Gaussian weight
$(3 \tau / \pi b^{2} t)^{3/2} \exp[-3  \tau (\bm{r} - \mu \bm{Q} + \mu'{\bm{Q}}')^{2} / b^{2} t]$.
The Fourier transform of this factor is simple:
\begin{equation}
 \label{foureir_transform_for_gaussian_factor_harmonic_dumbbell}
 \begin{split}
  & \int d\bm{r} \, e^{-i k r_{x}} 
  \left(\frac{3 \tau}{ \pi b^{2} t}\right)^{3/2} 
  \exp\left[ -\frac{3  \tau (\bm{r} - \mu \bm{Q} + \mu'{\bm{Q}}')^{2}}{b^{2} t}
      \right] \\
  & =   \left(\frac{3 \tau}{ \pi b^{2} t}\right)^{1/2} 
\int dr_{x} \, 
  \exp\left[ -\frac{3  \tau (r_{x} - \mu Q_{x} + \mu'Q_{x}')^{2}}{b^{2} t}
  - i r_{x} k
      \right] \\
  & = 
  \exp\left[ -\frac{b^{2} t}{12  \tau} k^{2}
  + i (- \mu Q_{x} + \mu'Q_{x}') k
      \right] .
 \end{split}
\end{equation}
From eqs~\eqref{shear_relaxation_modulus_harmonic_dumbbell}
and \eqref{foureir_transform_for_gaussian_factor_harmonic_dumbbell},
we have
\begin{equation}
 \label{shear_relaxation_modulus_harmonic_dumbbell_fourier_transform_modified}
 \begin{split}
  G(k,t) & = \frac{9 \nu k_{B} T}{b^{4}} 
\left[
     \frac{9}{4 \pi^{2} b^{4} (1 - e^{-2 t / \tau})}
    \right]^{3/2}   \int_{-1/2}^{1/2} d \mu  \int_{-1/2}^{1/2} d \mu' 
  \int d\bm{Q}d\bm{Q}' \, Q_{x} Q_{y} Q_{x}' Q_{y}' \\
  & \qquad \times 
  \exp\left[ 
   -\frac{b^{2} t}{12  \tau} k^{2}
  + i (- \mu Q_{x} + \mu'Q_{x}') k
  -  \frac{3 (\bm{Q}^{2} - 2 e^{-t / \tau} \bm{Q} \cdot \bm{Q}' + {\bm{Q}'}^{2})}{2 b^{2} (1 - e^{-2 t / \tau})}
  \right] .
 \end{split}
\end{equation}
The factor in the exponential function in eq~\eqref{shear_relaxation_modulus_harmonic_dumbbell_fourier_transform_modified}
can be rearranged as follows:
\begin{equation}
 \label{shear_relaxation_modulus_harmonic_dumbbell_fourier_transform_weight_factor}
  \begin{split}
   & i (- \mu Q_{x} + \mu'Q_{x}') k
  -  \frac{3 (\bm{Q}^{2} - 2 e^{-t / \tau} \bm{Q} \cdot \bm{Q}' + {\bm{Q}'}^{2})}{2 b^{2} (1 - e^{-2 t / \tau})} \\
   & = - \frac{3}{2 b^{2} (1 - e^{-2 t / \tau})}
    \left[ \bm{Q}
   - e^{-t / \tau} \bm{Q}' 
   + \frac{i \mu b^{2}(1 - e^{-2 t / \tau}) k \bm{e}_{x} }{3}   \right]^{2} \\
  & \qquad   - \frac{3}{2 b^{2}} \left[ {\bm{Q}'}^{2} 
  +  \frac{i b^{2} (\mu e^{-t / \tau}  - \mu' ) k \bm{e}_{x}}{3} \right]^{2}
  - \frac{b^{2}}{6} (\mu^{2} - 2 \mu \mu'  e^{-t / \tau}
    + {\mu'}^{2} ) k^{2}.
  \end{split}
\end{equation}
With eq~\eqref{shear_relaxation_modulus_harmonic_dumbbell_fourier_transform_weight_factor},
we can straightforwardly calculate the integrals
over $\bm{Q}$ and $\bm{Q}'$ in eq~\eqref{shear_relaxation_modulus_harmonic_dumbbell_fourier_transform_modified}.
We extract the part which depends on $\bm{Q}$ in
eq~\eqref{shear_relaxation_modulus_harmonic_dumbbell_fourier_transform_modified} and calculate it:
\begin{equation}
 \label{shear_relaxation_modulus_harmonic_dumbbell_integrals_q}
 \begin{split}
  & \left[
     \frac{3}{2 \pi b^{2} (1 - e^{-2 t / \tau})}
    \right]^{3/2} 
  \int d\bm{Q} \, Q_{x} Q_{y} \\
  & \qquad \times 
  \exp\left[ 
- \frac{3}{2 b^{2} (1 - e^{-2 t / \tau})}
    \left[ \bm{Q}
   - e^{-t / \tau} \bm{Q}' 
   + \frac{i \mu b^{2}(1 - e^{-2 t / \tau}) k \bm{e}_{x} }{3}   \right]^{2} 
  \right] \\
  & = e^{-t / \tau}
    \left[ e^{-t / \tau} Q'_{x}
   - \frac{i \mu b^{2}(1 - e^{-2 t / \tau}) k}{3}   \right]  Q'_{y}.
 \end{split}
\end{equation}
Then we extract the part which depends on $\bm{Q}'$ in eq~\eqref{shear_relaxation_modulus_harmonic_dumbbell_fourier_transform_modified}
and calculate it in a similar way.
\begin{equation}
 \label{shear_relaxation_modulus_harmonic_dumbbell_integrals_qp}
 \begin{split}
  & \left(
     \frac{3}{2 \pi b^{2}}
    \right)^{3/2} 
  \int d\bm{Q}' \, Q_{x}' \left[ e^{-t / \tau} Q'_{x}
   - \frac{i \mu b^{2}(1 - e^{-2 t / \tau}) k}{3}   \right] {Q'_{y}}^{2} \\
  & \qquad \times 
  \exp\left[ 
 - \frac{3}{2 b^{2}} \left[ {\bm{Q}'}^{2} 
  +  \frac{i b^{2} (\mu e^{-t / \tau}  - \mu' ) k \bm{e}_{x}}{3} \right]^{2}
  \right] \\
  & =  \frac{b^{2}}{3} \left(
     \frac{3}{2 \pi b^{2}}
    \right)^{1/2} 
  \int dQ_{x}' \, \left[ e^{-t / \tau} {Q'_{x}}^{2}
   - \frac{i \mu b^{2}(1 - e^{-2 t / \tau}) k}{3} Q_{x}'   \right]  
 \\
  & \qquad \times 
  \exp\left[ 
 - \frac{3}{2 b^{2}} \left[ {{Q}'_{x}}^{2} 
  +  \frac{i b^{2} (\mu e^{-t / \tau}  - \mu' ) k }{3} \right]^{2}
  \right] \\
  & =  \frac{b^{4}}{9} \left[ 
  e^{-t / \tau}
   + (\mu  - e^{-t / \tau} \mu') (\mu' -  e^{-t / \tau} \mu) \frac{b^{2} k^{2} }{3} \right]  .
 \end{split}
\end{equation}
Finally we can rewrite eq~\eqref{shear_relaxation_modulus_harmonic_dumbbell_fourier_transform_modified}
as eq~\eqref{shear_relaxation_modulus_harmonic_dumbbell_fourier_transform} in the main text.

Next we show the calculations for eq~\eqref{shear_relaxation_modulus_harmonic_dumbbell_fourier_transform_final}.
To calculate the integrals over $\mu$ and $\mu'$ in eq~\eqref{shear_relaxation_modulus_harmonic_dumbbell_fourier_transform},
we introduce the variable transform from $\mu$ and $\mu'$ to $\mu_{\pm} \equiv \mu \pm \mu'$.
The integrals over $\mu_{+}$ and $\mu_{-}$ is performed for a diamond-like region $\Omega$
which satisfies $|\mu_{+} + \mu_{-}| \le 1$ and $|\mu_{+} - \mu_{-}| \le 1$.
$\mu$ and $\mu'$ can be expressed by $\mu_{\pm}$ as $\mu = (\mu_{+} + \mu_{-}) / 2$
and $\mu' = (\mu_{+} - \mu_{-}) / 2$.
In eq~\eqref{shear_relaxation_modulus_harmonic_dumbbell_fourier_transform},
we have two factors which depend on $\mu$ and $\mu'$. Then cay be rewritten
in terms of $\mu_{\pm}$ as follows:
\begin{equation}
 \label{shear_relaxation_modulus_harmonic_dumbbell_mu_factor_first}
 \begin{split}
  & e^{- t / \tau} 
  + (\mu  - \mu' e^{-t / \tau}) (\mu' - \mu  e^{-t / \tau} ) \frac{b^{2} k^{2}}{3} \\
  & = e^{- t / \tau} 
  + \left[ (1 - e^{-t / \tau})^{2} \mu_{+}^{2}
  - (1 + e^{-t / \tau})^{2} \mu_{-}^{2}
  \right] \frac{b^{2} k^{2}}{12} \\
  & = 
   \frac{1}{2} \left[ 
   (1 + e^{-t / \tau}) (1 - \xi_{+} \mu_{-}^{2})
  - (1 - e^{-t / \tau}) (1 -  \xi_{-} \mu_{+}^{2})
  \right] ,
 \end{split}
\end{equation}
\begin{equation}
 \label{shear_relaxation_modulus_harmonic_dumbbell_mu_factor_second}
 \begin{split}
  &   - (\mu^{2} - 2 \mu \mu'  e^{-t / \tau}
    + {\mu'}^{2} ) \frac{b^{2} k^{2}}{6}\\
  & = - \left[ (1 - e^{-t / \tau})  \mu_{+}^{2}
  + (1 + e^{-t / \tau})  \mu_{-}^{2} \right] \frac{b^{2} k^{2}}{12} \\
  & = - \frac{1}{2} \xi_{-} \mu_{+}^{2} 
  - \frac{1}{2} \xi_{+} \mu_{-}^{2},
 \end{split}
\end{equation}
where we have defined $\xi_{\pm} \equiv b^{2} k^{2} (1 \pm e^{-t / \tau}) / 6$.
By using eqs~\eqref{shear_relaxation_modulus_harmonic_dumbbell_mu_factor_first}
and \eqref{shear_relaxation_modulus_harmonic_dumbbell_mu_factor_second} together with
$d\mu d\mu' = (1 / 2) d\mu_{+}d\mu_{-}$,
we can rewrite eq~\eqref{shear_relaxation_modulus_harmonic_dumbbell_fourier_transform}
as follows:
\begin{equation}
 \label{shear_relaxation_modulus_harmonic_dumbbell_fourier_transform_modified2}
 \begin{split}
  G(k,t) & = \frac{1}{4} \nu k_{B} T 
    \exp\left[ - 
  \left(1 + \frac{b^{2}k^{2}}{12}\right)\frac{t}{\tau}
  \right] \\
  & \qquad \times 
  \int_{\Omega} d \mu_{+}d\mu_{-} \, \left[ 
   (1 + e^{-t / \tau}) (1 - \xi_{+} \mu_{-}^{2})
  - (1 - e^{-t / \tau}) (1 -  \xi_{-} \mu_{+}^{2})
  \right]  \\
  & \qquad \times 
  \exp\left(
  - \frac{1}{2} \xi_{-} \mu_{+}^{2} 
  - \frac{1}{2} \xi_{+} \mu_{-}^{2}
  \right) \\
  & = \frac{1}{4} \nu k_{B} T 
    \exp\left[ - 
  \left(1 + \frac{b^{2}k^{2}}{12}\right)\frac{t}{\tau}
  \right]  \left[ 
   (1 + e^{-t / \tau}) I_{+}
  - (1 - e^{-t / \tau}) I_{-}
  \right] ,
\end{split}
\end{equation}
with $I_{\pm}$ defined as
\begin{equation}
 \label{shear_relaxation_modulus_harmonic_dumbbell_diamond_integrals}
 I_{\pm} \equiv 
  \int_{\Omega} d\mu_{+}d\mu_{-} \, (1 - \xi_{\pm} \mu_{\mp}^{2})
  \exp\left(
  - \frac{1}{2} \xi_{\mp} \mu_{\pm}^{2} 
  - \frac{1}{2} \xi_{\pm} \mu_{\mp}^{2}
  \right).
\end{equation}

We calculate $I_{\pm}$. The integrals over $\mu_{+}$ and $\mu_{-}$ in the region $\Omega$
can be rewritten as 
\begin{equation}
 \int_{\Omega} d\mu_{+}d\mu_{-} = \int_{-1}^{1} d\mu_{\pm} \int_{-1 + |\mu_{\pm}|}^{1 - |\mu_{\pm}|} d\mu_{\mp}.
\end{equation}
Then $I_{\pm}$ can be integrated over $\mu_{\mp}$ as
\begin{equation}
 \label{shear_relaxation_modulus_harmonic_dumbbell_diamond_integrals_modified}
 \begin{split}
 I_{\pm} 
  & =   \int_{- 1}^{1}d \mu_{\pm} \,
  \exp\left(
  - \frac{1}{2} \xi_{\mp} \mu_{\pm}^{2} 
  \right) \int_{-1 + |\mu_{\pm}|}^{1 - |\mu_{\pm}|} d \mu_{\mp} \,
   (1 - \xi_{\pm} \mu_{\mp}^{2})   \exp\left(
  - \frac{1}{2} \xi_{\pm} \mu_{\mp}^{2}
  \right) \\
  & = 2  \int_{- 1}^{1}d \mu_{\pm} \,
  (1 - |\mu_{\pm}|) \exp\left(
  - \frac{1}{2} \xi_{\mp} \mu_{\pm}^{2} 
  - \frac{1}{2} \xi_{\pm} (1 - |\mu_{\pm}|)^{2}
  \right) \\
  & = 4 \int_{0}^{1}d \mu_{\pm} \,
   (1 - \mu_{\pm}) 
  \exp\left[
  - \frac{1}{2} \xi_{\mp} \mu_{\pm}^{2} 
  - \frac{1}{2} \xi_{\pm} (1 - \mu_{\pm})^{2}
  \right] .
 \end{split}
\end{equation}
We calculate the integratal over $\mu_{\pm}$ in eq~\eqref{shear_relaxation_modulus_harmonic_dumbbell_diamond_integrals_modified}.
By rearranging the exponent in eq~\eqref{shear_relaxation_modulus_harmonic_dumbbell_diamond_integrals_modified},
we have
\begin{equation}
 \label{shear_relaxation_modulus_harmonic_dumbbell_diamond_integrals_modified2}
 \begin{split}
 I_{\pm} 
  & = 4 
  \exp\left[ - \frac{(1  - e^{-2t / \tau}) b^{2} k^{2}}{24}\right]
  \int_{0}^{1}d \mu_{\pm} \,
   (1 - \mu_{\pm}) 
  \exp\left[
  - \frac{b^{2} k^{2}}{6} \left( \mu_{\pm}
  - \frac{1 \pm e^{-t / \tau}}{2}
  \right)^{2}
  \right] \\
  & = 
  2 \exp\left[ - \frac{(1  - e^{-2t / \tau}) b^{2} k^{2}}{24}\right] 
  \left[ (1 \mp e^{-t / \tau})
  I' 
  - 2 I''_{\pm} \right],
 \end{split}
\end{equation}
with $I'$ and $I''_{\pm}$ defined as
\begin{align}
 I' & \equiv  \int_{0}^{1} d \mu \,
  \exp\left[
  - \frac{b^{2} k^{2}}{6} \left( \mu
  - \frac{1 \pm e^{-t / \tau}}{2}
  \right)^{2}
  \right], \\
 I_{\pm}'' & \equiv \int_{0}^{1}d \mu \,
   \left(\mu - \frac{1 \pm e^{-t / \tau}}{2} \right)
  \exp\left[
  - \frac{b^{2} k^{2}}{6} \left( \mu
  - \frac{1 \pm e^{-t / \tau}}{2}
  \right)^{2}
  \right].
\end{align}
When we calculate $G(k,t)$, $I'$ cancels and thus we do not need to calculate it further.
$I''_{\pm}$ can be calculated as
\begin{equation}
 \label{shear_relaxation_modulus_harmonic_dumbbell_diamond_integrals_part}
 \begin{split}
 I_{\pm}''
  & = \int_{0}^{1}d \mu \,
   \left(\mu - \frac{1 \pm e^{-t / \tau}}{2} \right)
  \exp\left[
  - \frac{b^{2} k^{2}}{6} \left( \mu
  - \frac{1 \pm e^{-t / \tau}}{2}
  \right)^{2}
  \right] \\
  & = \frac{3}{b^{2} k^{2}}
  \left[   \exp\left[
  - \frac{b^{2} k^{2}}{24} (1 \pm e^{-t / \tau})^{2}
  \right] -   \exp\left[
  - \frac{b^{2} k^{2}}{24} (1 \mp e^{-t / \tau})^{2}
  \right] \right] \\
  & = \mp \frac{6}{b^{2} k^{2}}
     \exp\left[
  - \frac{ (1 + e^{- 2t / \tau}) b^{2} k^{2}}{24}
  \right]
\sinh \left(
\frac{b^{2} k^{2}}{12}e^{-t / \tau} 
  \right) .
 \end{split}
\end{equation}
By combining eqs~\eqref{shear_relaxation_modulus_harmonic_dumbbell_fourier_transform_modified2},
\eqref{shear_relaxation_modulus_harmonic_dumbbell_diamond_integrals_modified2},
and \eqref{shear_relaxation_modulus_harmonic_dumbbell_diamond_integrals_part},
we have
\begin{equation}
 \label{shear_relaxation_modulus_harmonic_dumbbell_fourier_transform_modified3}
 \begin{split}
  G(k,t) 
 & =\nu k_{B} T 
    \exp\left[ - 
  \left(1 + \frac{b^{2}k^{2}}{12}\right)\frac{t}{\tau}
  - \frac{(1  - e^{-2t / \tau}) b^{2} k^{2}}{24}\right] \\
  & \qquad \times \left[ 
  -  (1 + e^{-t / \tau})  I''_{+} 
  + (1 - e^{-t / \tau}) I''_{-} 
  \right] \\
 & =\nu k_{B} T 
    \exp\left[ - 
  \left(1 + \frac{b^{2}k^{2}}{12}\right)\frac{t}{\tau}
  - \frac{ b^{2} k^{2}}{12} 
  \right]   \frac{12}{b^{2} k^{2}} \sinh \left(
\frac{b^{2} k^{2}}{12}e^{-t / \tau} 
  \right) .
\end{split}
\end{equation}
Finally we have eq~~\eqref{shear_relaxation_modulus_harmonic_dumbbell_fourier_transform_final}
in the main text.


\begin{thebibliography}{19}
\expandafter\ifx\csname natexlab\endcsname\relax\def\natexlab#1{}\fi
\expandafter\ifx\csname bibnamefont\endcsname\relax
  \def\bibnamefont#1{#1}\fi
\expandafter\ifx\csname bibfnamefont\endcsname\relax
  \def\bibfnamefont#1{#1}\fi
\expandafter\ifx\csname citenamefont\endcsname\relax
  \def\citenamefont#1{#1}\fi
\expandafter\ifx\csname url\endcsname\relax
  \def\url#1{\texttt{#1}}\fi
\expandafter\ifx\csname urlprefix\endcsname\relax\def\urlprefix{URL }\fi
\providecommand{\bibinfo}[2]{#2}
\providecommand{\eprint}[2][]{\url{#2}}

\bibitem{Waigh-2005}
\bibinfo{author}{\bibnamefont{Waigh} \bibfnamefont{TA}},
  \emph{\bibinfo{journal}{Rep. Prog. Phys.}},  \textbf{\bibinfo{volume}{68}},
  \bibinfo{pages}{685} (\bibinfo{year}{2005}).

\bibitem{Squires-Mason-2009}
\bibinfo{author}{\bibnamefont{Squires} \bibfnamefont{TM}},
  \bibinfo{author}{\bibnamefont{Mason} \bibfnamefont{TG}},
  \emph{\bibinfo{journal}{Annu. Rev. Fluid Mech.}},
  \textbf{\bibinfo{volume}{42}}, \bibinfo{pages}{413} (\bibinfo{year}{2009}).

\bibitem{Fox-1977}
\bibinfo{author}{\bibnamefont{Fox} \bibfnamefont{RF}},
  \emph{\bibinfo{journal}{J. Math. Phys.}},  \textbf{\bibinfo{volume}{18}},
  \bibinfo{pages}{2331} (\bibinfo{year}{1977}).

\bibitem{Evans-1981}
\bibinfo{author}{\bibnamefont{Evans} \bibfnamefont{DJ}},
  \emph{\bibinfo{journal}{Phys. Rev. A}},  \textbf{\bibinfo{volume}{23}},
  \bibinfo{pages}{2622} (\bibinfo{year}{1981}).

\bibitem{Alley-Alder-1983}
\bibinfo{author}{\bibnamefont{Alley} \bibfnamefont{WE}},
  \bibinfo{author}{\bibnamefont{Alder} \bibfnamefont{BJ}},
  \emph{\bibinfo{journal}{Phys. Rev. A}},  \textbf{\bibinfo{volume}{27}},
  \bibinfo{pages}{3158} (\bibinfo{year}{1983}).

\bibitem{Evans-Morris-book}
\bibinfo{author}{\bibnamefont{Evans} \bibfnamefont{DJ}},
  \bibinfo{author}{\bibnamefont{Morris} \bibfnamefont{GP}},
  \emph{\bibinfo{title}{``Statistical Mechanics of Nonequilibrium Liquids''}},
  \bibinfo{edition}{2nd}  ed.,
  (\bibinfo{year}{2008}), \bibinfo{publisher}{Cambridge University Press},
  \bibinfo{address}{Cambridge}.

\bibitem{Hansen-Daivis-Travis-Todd-2007}
\bibinfo{author}{\bibnamefont{Hansen} \bibfnamefont{JS}},
  \bibinfo{author}{\bibnamefont{Daivis} \bibfnamefont{PJ}},
  \bibinfo{author}{\bibnamefont{Travis} \bibfnamefont{KP}},
  \bibinfo{author}{\bibnamefont{Todd} \bibfnamefont{BD}},
  \emph{\bibinfo{journal}{Phys. Rev. E}},  \textbf{\bibinfo{volume}{76}},
  \bibinfo{pages}{041121} (\bibinfo{year}{2007}).

\bibitem{Glavatskiy-Dalton-Daivis-Todd-2015}
\bibinfo{author}{\bibnamefont{Glavatskiy} \bibfnamefont{KS}},
  \bibinfo{author}{\bibnamefont{Dalton} \bibfnamefont{BA}},
  \bibinfo{author}{\bibnamefont{Daivis} \bibfnamefont{PJ}},
  \bibinfo{author}{\bibnamefont{Todd} \bibfnamefont{BD}},
  \emph{\bibinfo{journal}{Phys. Rev. E}},  \textbf{\bibinfo{volume}{91}},
  \bibinfo{pages}{062132} (\bibinfo{year}{2015}).

\bibitem{Doi-Edwards-book}
\bibinfo{author}{\bibnamefont{Doi} \bibfnamefont{M}},
  \bibinfo{author}{\bibnamefont{Edwards} \bibfnamefont{SF}},
  \emph{\bibinfo{title}{``The Theory of Polymer Dynamics''}},
  (\bibinfo{year}{1986}), \bibinfo{publisher}{Oxford University Press}, \bibinfo{address}{Oxford}.

\bibitem{Irving-Kirkwood-1950}
\bibinfo{author}{\bibnamefont{Irving} \bibfnamefont{JH}},
  \bibinfo{author}{\bibnamefont{Kirkwood} \bibfnamefont{JG}},
  \emph{\bibinfo{journal}{J. Chem. Phys.}},  \textbf{\bibinfo{volume}{18}}
  (\bibinfo{year}{1950}).

\bibitem{Schfield-Henderson-1982}
\bibinfo{author}{\bibnamefont{Schofield} \bibfnamefont{P}},
  \bibinfo{author}{\bibnamefont{Henderson} \bibfnamefont{JR}},
  \emph{\bibinfo{journal}{Proc. Roy. Soc. Lond. A: Math. Phys.}},
  \textbf{\bibinfo{volume}{379}} (\bibinfo{year}{1982}).

\bibitem{Landau-Lifshitz-book}
\bibinfo{author}{\bibnamefont{Landau} \bibfnamefont{LD}},
  \bibinfo{author}{\bibnamefont{Lifshitz} \bibfnamefont{EM}},
  \emph{\bibinfo{title}{``Theory of Elasticity''}}, \bibinfo{edition}{3rd} ed.,
  (\bibinfo{year}{1986}), 
  \bibinfo{publisher}{Butterworth-Heinemann}, \bibinfo{address}{Oxford}.

\bibitem{Uneyama-Nakai-Masubuchi-2019}
\bibinfo{author}{\bibnamefont{Uneyama} \bibfnamefont{T}},
  \bibinfo{author}{\bibnamefont{Nakai} \bibfnamefont{F}},
  \bibinfo{author}{\bibnamefont{Masubuchi} \bibfnamefont{Y}},
  \emph{\bibinfo{journal}{Nihon Reoroji Gakkaishi (J. Soc. Rheol. Jpn.)}},
  \textbf{\bibinfo{volume}{47}}, \bibinfo{pages}{143} (\bibinfo{year}{2019}).

\bibitem{Gardiner-book}
\bibinfo{author}{\bibnamefont{Gardiner} \bibfnamefont{CW}},
  \emph{\bibinfo{title}{``Handbook of Stochastic Methods''}},
  \bibinfo{edition}{3rd} ed.,
  (\bibinfo{year}{2004}), 
  \bibinfo{publisher}{Springer}, \bibinfo{address}{Berlin}.

\bibitem{Ewen-Richter-1997}
\bibinfo{author}{\bibnamefont{Ewen} \bibfnamefont{B}},
  \bibinfo{author}{\bibnamefont{Richter} \bibfnamefont{D}},
  \emph{\bibinfo{journal}{Adv. Polym. Sci.}},  \textbf{\bibinfo{volume}{134}},
  \bibinfo{pages}{1} (\bibinfo{year}{1997}).

\bibitem{Ottinger-book}
\bibinfo{author}{\bibnamefont{{\"O}ttinger} \bibfnamefont{HC}},
  \emph{\bibinfo{title}{``Stochastic Processes in Polymeric Fluids: Tools and
  Examples for Developing Simulation Algorithms''}},
  (\bibinfo{year}{1996}), \bibinfo{publisher}{Springer}, Berlin.

\bibitem{Kroger-2004}
\bibinfo{author}{\bibnamefont{Kr\"{o}ger} \bibfnamefont{M}},
  \emph{\bibinfo{journal}{Phys. Rep.}},  \textbf{\bibinfo{volume}{390}},
  \bibinfo{pages}{453} (\bibinfo{year}{2004}).

\bibitem{vanKampen-book}
\bibinfo{author}{\bibnamefont{van Kampen} \bibfnamefont{NG}},
  \emph{\bibinfo{title}{``Stochastic Processes in Physics and Chemistry''}},
  \bibinfo{edition}{3rd} ed.,
   (\bibinfo{year}{2007}),
  \bibinfo{publisher}{Elsevier}, \bibinfo{address}{Amsterdam}.

\bibitem{Ooura-Mori-1991}
\bibinfo{author}{\bibnamefont{Ooura} \bibfnamefont{T}},
  \bibinfo{author}{\bibnamefont{Mori} \bibfnamefont{M}},
  \emph{\bibinfo{journal}{J. Comp. Appl. Math.}},
  \textbf{\bibinfo{volume}{38}}, \bibinfo{pages}{353} (\bibinfo{year}{1991}),
  \bibinfo{note}{https://www.kurims.kyoto-u.ac.jp/\~{}ooura/intde.html}.

\end{thebibliography}

\end{document}